\documentclass[12pt]{iopart}

\usepackage{iopams} 
\usepackage{setstack}
\usepackage{graphicx}

\usepackage{soul}
\usepackage{hyperref, xcolor}
\hypersetup{
    colorlinks=true,
    linkcolor=[rgb]{0 0 0.5},
    urlcolor=cyan,
    citecolor= [rgb]{0 0 0.5}
}
\renewcommand{\comment}[1]{}

\renewcommand{\bs}{\boldsymbol}

\renewcommand{\k}{{\bs k}}
\renewcommand{\v}{{\bs v}}
\renewcommand{\o}{\bs{\Omega}}
\newcommand{\uo}{\underline{\Omega}}
\newcommand{\ez}{\bs{e}_z}
\renewcommand{\bxi}{\bs{\xi}}
\newcommand{\txi}{\tilde{\xi}}

\newcommand{\R}{\bs{\mathcal{R}}}

\renewcommand{\r}{{\bs r}}
\newcommand{\p}{\bs{p}}

\newcommand{\cz}{{\rm r}} 
\newcommand{\psv}{\zeta} 
\renewcommand{\d}{d}

\newcommand{\brI}{{\mathbf I}}

\newcommand{\rD}{{\rm D}}

\newcommand{\la}{\left\langle}
\newcommand{\ra}{\right\rangle}

\renewcommand{\d}{{\rm d}}
\newcommand{\ee}{{\rm e}}
\newcommand{\ii}{{\rm i}}
\newcommand{\TT}{{\rm T}}

\newcommand{\ozg}{\vartheta}

\newcommand{\BvL}{Bohr-Van Leeuwen}
\newcommand{\FP}{Fokker-Planck}

\newcommand{\BEA}{\begin{eqnarray}}
\newcommand{\EEA}{\end{eqnarray}}

\newcommand{\rx}[0]{{\rm x}}
\newcommand{\brx}{\mathbf{x}}

\renewcommand{\text}[1]{{\rm #1}}
\newcommand{\eqref}[1]{(\ref{#1})}
\newcommand{\tfrac}{\case}

\begin{document}

\title{Weak (non)conservation and stochastic dynamics of angular momentum}

\author{Ashot Matevosyan$^{1,2}$}

\address{$^1$ Max Planck Institute for the Physics of Complex Systems, N\"{o}thnitzerstra{\ss}e 38, Dresden, 01187, Germany}
\address{$^2$ Alikhanyan National Laboratory (Yerevan Physics Institute), 2 Alikhanyan Brothers street, Yerevan, 0036, Armenia}

\ead{ashmat@pks.mpg.de}
\vspace{10pt}
\begin{indented}
\item[]ORCID: 0000-0002-6628-6676
\end{indented}

\begin{abstract}
Angular momentum conservation influences equilibrium statistical mechanics, leading to a generalized microcanonical density for an isolated system and a generalized Gibbs density for a weakly coupled system. We study the stochastic decay of angular momentum due to weakly imperfect rotational symmetry of the external potential that confines the isolated many-particle system. We present a mesoscopic description of the system, deriving Langevin and Fokker-Planck equations, which are consistent with equilibrium statistical mechanics when rotational symmetry is maintained. When the symmetry is weakly violated, we formulate a coarse-grained stochastic differential equation governing the decay of total angular momentum over time. To validate our analytical predictions, we conduct numerical simulations of the microcanonical ensemble, an isolated system undergoing thermalization due to weak two-body interactions. Our coarse-grained Langevin equation accurately characterizes both the decay of the angular momentum and its fluctuations in a steady state. Furthermore, we estimate the parameters of our mesoscopic model directly from simulations, providing insights into the dissipative phenomenological coefficients, such as friction. More generally, this study contributes to a deeper understanding of the behavior of the integrals of motion when the corresponding symmetry is weakly violated.
\end{abstract}
%
\vspace{2pc}
\noindent{\it Keywords}: 
angular momentum, 
integrals of motion, 
Langevin,
Fokker-Planck,
rotational symmetry,
violated symmetry
%
%
%
%

\section{Introduction}

Rotational symmetry conserves the angular momentum of an isolated multi-particle system. The angular momentum is an additive quantity and its conservation changes the equilibrium statistical mechanics of the system, which is now described via the modified microcanonical density that, besides energy conservation, also ensures the angular momentum conservation. Likewise, the equilibrium density of a weakly-coupled subsystem changes toward the generalized Gibbs distribution. These facts are well-known \cite{landau_stat}. The rotation symmetry, however, is never perfect and only approximate. Consequently, the orbital momentum gradually decays over time, dissipating the energy of the ordered rotation and increasing the system's temperature. I would like to describe this process both analytically and numerically and understand its mechanisms. 

The first motivation for studying this problem comes from microphysical situations that support a rotatory motion in quasi-equilibrium states. Rotary motors play an important role in storing and transmitting energy during the functioning of molecular machines
\cite{venturi}. The best known natural rotary motor is the ATP synthase enzyme that manufactures ATP, the basic battery of biological
energy \cite{venturi,splendid, marvellous,mugnaiRMP}. Here, rotation pertains to the dynamic movement of specific molecular components, also known as the rotor, within the enzyme complex ATPase. This protein complex imposes a confining potential on the rotor, which is not perfectly symmetric. This will contribute to the dissipation of the angular momentum of the rotor. Another interesting example of rotatory equilibrium motion is provided by rotating atomic clusters \cite{clusters}. Yet another class of rotating equilibrium systems is provided by classical Brownian charges, where under the influence of a constant magnetic field the phonon bath acquires an orbital momentum \cite{lasting}. Rotatory motion is also present in the opposite realm of very large systems, i.e. self-gravitating star clusters \cite{laliena}.

We opine that studying the above statistical physics problem will eventually clarify the turbulent hydrodynamic situation, where large vortices fraction into smaller ones. This decay process is primarily driven by viscous forces acting around the vortex \cite{kolmogorovlegacy,turbulenceroselessons,turbulencerose}, and results in the well-known energy cascade. As a side note, this research asks the same question (what happens to the integrals of motion when the system is weakly perturbed?) in the context of ergodic systems, as the known Kolmogorov-Arnold-Moser theorem does for integrable systems \cite{kolmogorov1954conservation, KAMrangarajan}.

We examine an isolated system of interacting point-like particles confined within a time-independent potential. 
Rotations around the $z$-axis leave the potential energy invariant. Time-translation invariance and rotation invariance imply two fundamental properties: conservation of the total energy and conservation of the $z$ component of the total angular momentum. Hence, the microcanonical equilibrium density for this system reads \cite{landau_stat}
\begin{eqnarray}\label{microcanonical}
    \rho_\text{m}(\psv)\propto\delta\left(E-H(\psv)\right)\delta\left(L_z-L_z(\psv)\right),
\end{eqnarray}
where $\psv$ represents phase-space variables, e.g. positions and momenta of particles, $H$ is the Hamiltonian and $L_z$ is the $z$ component of the total angular momentum. On the other hand, smaller subsystems, which may consist of just a single particle, in the thermodynamic limit, can be described by the canonical distribution function \cite{terletski, berdichevsky, touchette}: 
\begin{equation}\label{canonic}
    \rho_\text{c}(\psv_\text{sub})\propto \exp\left(-\beta \,H_{\rm sub}(\psv_{\rm sub})-\alpha \, L_{z,{\rm sub}}(\psv_{\rm sub}) \right),
\end{equation}
where $H_{\rm sub}$ and $L_{z,{\rm sub}}$ represent the energy and the $z$ component of angular momentum of the subsystem, and $\psv_{\rm sub}$ denotes its phase-space coordinates.
The parameters ${\alpha}$ and $\beta$ characterize the thermal bath, i.e.
the rest of the system, and are the same for all (small) subsystems. 
On the other hand, entropy maximization leads to a uniform temperature and uniform rotation with angular velocity $\o  = \bs{e}_z \Omega_z$ \cite{landau_stat}($\bs{e}_z$ is the unit vector in $z$ direction). 
In other words, in equilibrium, the system macroscopically rotates as a bulk.
Hence, $\beta=T^{-1}$ represents the inverse temperature, and ${\alpha}=-\beta \,\Omega_z$
\footnote{This can be seen by considering a subsystem with Hamiltonian \eqref{subsystem-hamiltonian} at specific location $\r$, \eqref{canonic} conditioned at $\r$ gives average velocity $\la \v \ra_{\rho_\text{c}} = -(\alpha/\beta) \bs{e}_z \times \r =\Omega_z \bs{e}_z \times \r$.
Also, $-\frac{\alpha}{\beta}=\frac{\langle x v_y-y v_x\rangle_{\rho_\text{c}}}{\langle x^2 + y^2 \rangle_{\rho_\text{c}}} = \frac{\la L \ra }{\la I \ra} = \Omega_z $. See Section \ref{sec-micro} for definitions.}. 
Distribution (\ref{canonic}) is an example of a generalized Gibbs distribution; such a distribution has attracted attention also for quantum systems \cite{gengibbs1,gengibbs2,gengibbs3}.

Equations (\ref{microcanonical}, \ref{canonic}) are two complementary thermodynamic descriptions of the system, applicable \emph{only} when the system exhibits rotational symmetry. When the rotational symmetry is broken, the angular momentum must decay to zero. This is due to the fact that the microcanonical distribution now only constrains the energy, expressed as $\rho_\text{m}(\psv)\propto\delta\left(E-H(\psv)\right)$, and the average total angular momentum with respect to this distribution becomes zero \footnote{just because the Hamiltonian is even function of momentum $\p$, while angular momentum $\r\times\p$ is odd function of $\p$, and therefore, average vanishes.} 

We study the case when the angular momentum is weakly conserved, such as when the external potential has a slight anisotropy.
The degree of anisotropy of the external potential determines the rate at which angular momentum decays.
Thus, the decay timescale can be significantly longer than the \emph{local relaxation time} of the system, by which we mean the relaxation (or thermalization) time of the isotropic system. This results in timescale separation and enables us to treat temperature and angular velocity as slow variables, representing a form of adiabatic assumption.

To derive the dynamics of these slow variables, we establish a mesoscopic description of a subsystem of our system, i.e. of a single particle. (We say ``mesoscopic'' because microscopic description is reserved for the Newtonian dynamics of single particles, as employed in molecular dynamics; see below.) 
To this end, we make a natural assumption that for zero angular momentum, the subsystem is described by the ordinary Langevin equation with a constant (i.e. time and space-independent) friction constant. From that assumption, we derive the Langevin (and then the Fokker-Planck) equation for a subsystem coupled to a thermal bath at temperature $T$, and with a non-zero angular momentum. The latter enters into the Langevin equation via the angular velocity $\Omega_z$; see section 2 for details. If the rotation symmetry is maintained, \eqref{canonic} serves as the steady-state solution to the \FP{} equation. Once $T$ and $\Omega_z$ change slowly, we consider the quasi-stationary solution of the \FP{} equation and employ it to find a coarse-grained stochastic equation that describes the decay of the total angular momentum. The overall temperature increases during this decay, as the kinetic energy associated with ordered motion (rotation) undergoes an irreversible transformation into the energy of chaotic motion. 

To substantiate our analytical findings, we conducted numerical simulations of the microcanonical ensemble, an isolated system that relaxes due to weak interactions. The application of the Langevin and \FP{} equations to a subsystem of an isolated equilibrating system is subject to certain conditions \cite{oppenheim-uses-abuses,mazur-molecular-theory,van1986brownian}, and sometimes the white-noise Markovian Langevin equation is not enough to fit the data \cite{langevingeneralized}. In general, it is rare to find studies that check the applicability of the Langevin equation to systems with many-body interactions. Moreover, we apply a modified Langevin equation describing a rotating system, and we are not aware of works studying the applicability of the Langevin equations in this case. We demonstrate that, despite the simplicity of the model, the phenomenological Langevin equations indeed agree with simulations. The stochastic equation derived from our analysis precisely characterizes the decay of the total angular momentum, as well as its fluctuations. Moreover, we assess the dissipative phenomenological coefficient (friction) by examining both the dissipation of total angular momentum and its fluctuations in the steady state. In numerical simulations, we had to consider a dilute, low-density system, so that interactions were weak and, therefore, the canonical and microcanonical distributions were consistent.

This paper is organized as follows. In the next section, we derive the Langevin and \FP{} equations as a mesoscopic description for our system.
Then, in Section \ref{sec-L}, we go through the derivation of the stochastic evolution of the total angular momentum in the case of violated symmetry. In Section \ref{sec-stationary}, we list the conditions for the equivalence of microcanonical \eqref{microcanonical} and canonical \eqref{canonic} distributions and establish a connection between their parameters. The comparisons with numerical simulations are presented in Section \ref{sec-numerics}. The final section summarizes and provides perspectives on future research.

We placed the detailed derivations of the results and information about numerical analysis in the appendices, ensuring that the conclusions of this work are self-contained. \ref{app-derivation} offers an alternative perspective on the derivation of the \FP{} and Langevin equations. In \ref{app-FP-OU} we provide a general solution of \FP{} equation corresponding to the Ornstein-Uhlenbeck process. 
Autocorrelation functions are derived in \ref{app-autocor}. \ref{app-TO} and \ref{app-gammafit} elaborate on the estimation of the quantities from simulations. In \ref{app-micro-avg}, we show that a locally-equilibrium (or frozen) microcanonic ensemble does not explain the decay of angular momentum. \ref{app-space-gamma} considers a non-uniform friction constant as a possible generalization of the model. The possibility of collisionless relaxation is discussed in \ref{app-collisionless}. For in-depth details about simulations and figures, please refer to \ref{app-numerics}.

\section{Mesoscopic description of the system}\label{sec-micro}

Let us begin by considering the ergodic microcanonical ensemble $\delta(E-H(\psv))$, where only the total energy is conserved. The corresponding canonical distribution takes the form of the Gibbs distribution $\exp(-\beta\,H_\text{sub}(\psv_\text{sub}))$ where $\psv_\text{sub}$ denote the phase-space coordinates of the subsystem, and $\beta=T^{-1}$ is the inverse temperature; cf.~(\ref{canonic}). 

The Langevin equation is a standard tool for describing the dynamics of subsystems in statistical physics. It has been deduced from various phenomenological and/or microscopic approaches, and its validity limits are now well understood \cite{landau_stat,terletski,risken,langevingeneralized,de2013non,mazur-molecular-theory,oppenheim-uses-abuses}. Hence, we postulate that a single particle of the closed system is described via the following white-noise Langevin equation \cite{risken}:
\begin{eqnarray}\label{langevin-straight}
    \tfrac{\d}{\d t} \r = \v,
    \quad
    \tfrac{\d}{\d t} \v = 
    -\nabla_{\r} V(\r)
    -\gamma \v
    +\sqrt{2\gamma T} \;\bxi(t), \quad 
    \\
    \langle \bxi_i(t)\bxi_j(t')\rangle=\delta_{ij}\,\delta(t-t'),
\end{eqnarray}
where a single particle with a unit mass moves in a medium of temperature $T$. Here $\r=(x,y,z)$ and $\v=(v_x,v_y,v_z)$ are the position and velocity of the particle, while $V(\r)$ denotes the external, confining potential. $\nabla_{\r}=({\partial_x},\partial_y,\partial_z)$ denotes gradient by position. The interaction of the particle with the rest of the system is broken into two components: the friction force $-\gamma \v$ (deterministic part) and a white Gaussian noise $\sqrt{2\gamma T} \,\bxi(t)$. This is an isotropic and stochastic force with $\delta_{ij}$ being the Kronecker delta in (\ref{langevin-straight}). The noise amplitude and friction constant satisfy the Fluctuation-Dissipation relation \cite{risken}. Consequently, the stationary distribution of this process is indeed the canonical distribution mentioned above,  with the subsystem Hamiltonian \cite{risken}:
\begin{eqnarray}\label{subsystem-hamiltonian}
    H_\text{sub}(\psv_\text{sub}) = \frac{\v^2}{2} + V(\r)
    \qquad \psv_\text{sub} = (\r, \v)
\end{eqnarray}

We now derive a Langevin equation to model the motion of a particle within a rotating environment with angular velocity $\o=\hat{\bs{z}}\,\Omega$ and temperature $T$. Consider a small neighborhood surrounding the particle. Let the neighborhood be large enough to be considered to be in local equilibrium, yet small enough to assume a uniform average velocity $\o\!\times\! \r$ inside it ( $\times$ denotes cross-product between 3d vectors). 
Therefore, the relative velocity of a particle with respect to this neighborhood is $\v - \o\!\times\! \r$.
We assume the friction force to be proportional to the relative velocity instead of $\v$.
Furthermore, as the surroundings are in local equilibrium, the particle experiences an isotropic thermal noise of temperature $T$, as in \eqref{langevin-straight}. To account for this behavior, we propose a modification of the friction term in equation \eqref{langevin-straight} as follows:
\begin{eqnarray}
    \label{langevin}
    &\tfrac{\d}{\d t} \r = \v,
    \qquad
    \tfrac{\d}{\d t} \v = 
    -\nabla_{\r} V(\r)
    -\gamma \left(\v - \o\!\times\! \r\right)
    +\sqrt{2\gamma T} \;\bxi(t).
\end{eqnarray}    
A similar Langevin equation is used in Ref.~\cite{plasma-magnetizing}.
In \ref{app-derivation}, we derive the same Langevin equation \eqref{langevin} by changing to a rotating frame of reference, where the thermostat is at rest. By accounting for all the inertial forces, we write the usual Langevin equation \eqref{langevin-straight} in the rotating frame, which transforms into \eqref{langevin} in the rest (non-rotating) frame.

We write the \FP{} equation (also called the Kramers equation) corresponding to the above Langevin using standard procedures\cite{risken}
\begin{eqnarray}
    \label{FP0}
    \tfrac{\partial}{\partial t} P=& 
    -\nabla_{\r} \left[ \v P \right]+\nabla_{\v}  \left[ \left(\nabla_{\r} V(\r)
    +\gamma (\v - \o\!\times\! \r) \right) P \right]+ \gamma T \,\nabla_{\v}^2 P,
\end{eqnarray}
where $P=P(t, \r, \v)$ is the distribution function over $\r$ and $\v$ at time $t$. $\nabla_{\r}=({\partial_x},\partial_y,\partial_z)$ and $\nabla_{\v}=(\partial_{v_x},\partial_{v_y},\partial_{v_z})$ are gradient by position and velocity, respectively.
It can be shown (e.g. by substitution) that canonical distribution in \eqref{canonic} is a stationary solution of the \FP{} equation \eqref{FP0} \emph{if} $V(\r)$ is rotation symmetric. This is a consistency check for our analysis. From now on, we will employ \eqref{langevin} as our mesoscopic model. It has a characteristic relaxation time denoted as $\tau_r$, which we refer to as the \emph{local relaxation time}.

We emphasize here two limitations of this model. The friction coefficient $\gamma$ in \eqref{langevin} is a phenomenological coefficient, which is frequently taken as a constant \cite{risken}. We shall mostly follow this practice. However, more realistically, $\gamma$ is expected to depend on temperature. Moreover, it is an increasing function of temperature, which is intuitively expected because larger temperatures bring more frequent inter-particle collisions, which are behind the Langevin equation (\ref{langevin}). This intuition is confirmed e.g. for a temperature-dependent viscosity of the Lennard-Jones gas \cite{chapman-cowling}. In our case, as the total angular momentum decays, the kinetic energy of rotational motion irreversibly transforms into kinetic energy of chaotic motion, and thus the temperature increases. Hence, $\gamma$ is expected to increase during the decay of the angular momentum. Another limitation of using a constant $\gamma$ is that the particle density is not uniform, hence one expects a density-dependent friction constant. In \ref{app-space-gamma} we extend the analysis of the next section for space-dependent $\gamma(\r)$.

\section{Weak non-conservation of angular momentum}
\label{sec-L}

The previous section focused on a subsystem of a many-particle system. In this section, we return to the perspective of the closed system: we consider $N$ interacting particles with radius vectors $\{\r_i=(x_i,y_i,z_i) \}_{i=1}^N$ that interact with each other via translationally and rotationally invariant pairwise potentials $\sum_{i<j=1}^N {\cal U}(|\r_i-\r_j|)$, and move in external potential 
\begin{equation}\label{potential-perturbed}
    V(\r) = \frac{1}{2}\left(a^2(1+\epsilon)^2 \; x^2 + a^2(1-\epsilon)^2 \;
y^2 + a_z^2 \;z^2\right).
\end{equation}
The interaction potential must satisfy several conditions (e.g. short-range and weak interactions) for the equivalence of the canonical and microcanonical ensemble;, we discuss this in Section \ref{sec-stationary}.
Note that even with harmonic potential \eqref{potential-perturbed}, Langevin equation \eqref{langevin} in the $x$ and $y$ directions do not decouple. When the system rotates with $\bs{\Omega}=\ez \Omega$, the modified friction term couples $x$ and $y$ motion.

In (\ref{potential-perturbed}), the rotation symmetry around the $z$-axes is weakly broken due to $\epsilon\not=0$. Below, we still calculate the angular momentum and torques with respect to the $z$ axis.
As a result, the $z$ component of angular momentum, $L_z\equiv L$ is no longer conserved. More precisely,  
starting from the Newton equations with potentials $\sum_{i=1}^NV(\r_i)+\sum_{i<j=1}^N {\cal U}(|\r_i-\r_j|)$ for the $N$-particle system, one derives that $\frac{\d}{\d t}L$ is equal to the torques from external potential only:
\begin{eqnarray}
\tfrac{\d}{\d t}L\equiv    \tfrac{\d}{\d t}L_z = \left[\sum_{i=1}^N \r_i \times (-\nabla_{\r_i} V(\r_i))\right]_z
    = \sum_{i=1}^N 4 a^2 \epsilon \; x_i y_i
    \label{dL-sum}
\end{eqnarray}
where we note that the inter-particle interactions described by ${\cal U}$ 
cancel out from (\ref{dL-sum}) due to their rotation invariance. We shall now underline our main assumption, which is confirmed below (also numerically). We assume that there is a timescale separation: $L$ is a slow variable, and the microscopic degrees of freedom relax to a quasi-stationary state for a fixed $L$. This state is to be determined from (\ref{FP0}, \ref{potential-perturbed}). Given this assumption, we can separate $\sum_i^N 4 a^2 \epsilon \; x_i y_i$ into two terms: the time average over fast times (i.e. times that are larger than the relaxation time to the quasi-stationary state, but smaller than the characteristic time of changing $L$), and the difference between the actual $\sum_i^N 4 a^2 \epsilon \; x_i y_i$ and its time-average, which we shall eventually approximate as a random noise and calculate its features from (\ref{langevin}, \ref{potential-perturbed}). Hence we get from (\ref{dL-sum}):
\begin{eqnarray}
\label{dL}
\tfrac{\d}{\d t}L=\,4 a^2 \epsilon \,N\,\la xy\ra + 4 a^2 \epsilon \,\sqrt{N}\,h(t),
\end{eqnarray}
where $\la xy\ra$ is the time average at time $t$ and  $h(t)$ is defined as
\begin{eqnarray}\label{h}
    h(t)=\frac{1}{\sqrt{N}}\sum_{i=1}^{N}h_i(t),
    \quad h_i(t)=x_i(t)y_i(t)-\la
x y\ra_t.
\end{eqnarray}
Here $\la h(t) \ra = \la h_i(t) \ra = 0$ and $\la h(t) \ra = \la h_i(t) \ra \equiv \sigma_t^2$. Although we do not explicitly obtain the distribution of $h_i$, by the Central Limit Theorem, for a large number of particles $N$, the distribution of $h(t)$ approaches the Gaussian. 

First, we consider the deterministic part in \eqref{dL}. When $\epsilon=0$, the stationary distribution $\rho_\text{c}$ is described by equation \eqref{canonic}, and as a result, $\la x y \ra_{\rho_\text{c}}=0$. 
Therefore, the decay of $L$  is attributed to two effects: the perturbation of the external potential and the perturbation of the distribution function $\rho_\text{c}$, leading to a non-zero $\la xy \ra$ in (\ref{dL}). These two contributions are multiplied in \eqref{dL}, suggesting that the non-zero deterministic part of $\tfrac{\d}{\d t}L$ 
is a second-order effect over $\epsilon$.
Let $\tau_L$ be the characteristic time of the decay of $L$, which, as we discussed above, is proportional to  $\epsilon^{-2}$. It can be made arbitrarily long by choosing a sufficiently small perturbation $\epsilon$.

The distribution function $P(t, \r, \v)$ in \eqref{FP0} evolves over a timescale $\tau_L$, and we suppose it is much larger compared to the characteristic relaxation time (local relaxation time) $\tau_r$ of the \FP{} equation. This allows us to neglect the time derivative in the \FP{} and consider its quasi-stationary distribution. 
Moreover, the ergodicity of the system implies the equivalence of the time average and ensemble average in the (quasi)stationary state. Therefore $\la\;\ra$ can be interpreted as the average with respect to quasi-stationary distribution.

The $\la xy\ra$ is a single-particle property, and the single-particle distribution is governed by Langevin \eqref{langevin} and \FP{} \eqref{FP0}. 
With the quadratic potential defined in \eqref{potential-perturbed}, the \FP{} \eqref{FP0} becomes exactly solvable after dropping the time dependence.
We present the complete solution in \ref{app-FP-OU}, but in this section, we are only interested in the following 
(see \eqref{xy-mean-app} in \ref{app-FP-OU}) 
\begin{eqnarray}\label{xy-mean}
    \la x y\ra &=& -\frac{2\epsilon T a^2 \Omega \left(a^2+\gamma^2\right) }{\gamma(a^2-\Omega^2)\left(a^4+\gamma^2 
\Omega^2\right)}+ {\cal O}(\epsilon^2)
\end{eqnarray}
Note that this is non-zero because of $\epsilon$ perturbation. It is important to mention that the quasi-stationary distribution does not follow from a naive microcanonical distribution \eqref{microcanonical} with symmetric potential replaced by anisotropic potential \eqref{potential-perturbed}, see \ref{app-micro-avg}.

In \eqref{xy-mean}, $\la x y\ra$ explicitly depends on time through the time dependence of the slowly varying variables $T$ and $\Omega$.
Substituting this into \eqref{dL}, we get
\begin{eqnarray}\label{dl-mean}
    \la \tfrac{\d}{\d t}L \ra &= -
     \frac{4a^2 \epsilon^2}{\gamma} \,\frac{2\Omega N T}{a^2 - \Omega ^2}\;\frac{a^2(a^2+\gamma^2)}{a^4+\gamma^2\Omega^2}
\end{eqnarray}
This defines the characteristic timescale for the decay of the angular momentum: $\tau_L =\frac{\gamma}{4a^2\epsilon^2}$. Note that $L$ always decays regardless of the sign of $\epsilon$.
Note that the right-hand side depends on the temperature $T$ and angular velocity $\Omega$. In the following section, we establish their relationship with the angular momentum $L$ and thus close the equation \eqref{dl-mean}.

To characterize the fluctuations, we examine the autocorrelation function $C_h(t)$ of a process $h_i(t)$ (recall \eqref{h} for the definition):
\begin{eqnarray}
    C_h(t-t') = \la h_i(t) h_i(t')\ra = \la x_i(t)y_i(t)\; x_i(t')y_i(t')\ra - \la x_i y_i\ra^2,
\end{eqnarray}
From symmetry, this is the same for all particles and does not depend on the index $i$.
In a quasi-stationary state, correlation functions depend on the time difference $t-t'$ only. 

$C_h(t)$ is calculated in \ref{app-autocor} using the autocorrelation functions of Langevin equation \eqref{langevin} and using the formula of higher moments of Gaussian distribution. On the other hand, the autocorrelation function of process $h(t)$ reads
\begin{eqnarray}
    \la h(t)h(t')\ra &=& \frac{1}{N}\sum_{i=1}^N \la h_i(t)h_i(t)\ra + \frac{1}{N}\sum_{i\ne j=1}^N \la h_i(t)h_j(t)\ra
    \\
    &=& C_h(t-t') + \text{(cross\;terms)}
\end{eqnarray}
The cross-terms come from weak correlations between processes $\{h_i(t)\}_{i=1}^N$. However, the Central Limit Theorem implies that for sufficiently large $N$, the contribution from cross terms is negligible. Therefore, the autocorrelation function of $h(t)$ is the same $C_h(t)$ for the process $h_i(t)$ above.
$C_h(t)$ oscillates at frequency $2a$, with the amplitude of the oscillations decaying over a timescale
\begin{eqnarray}\label{tau-h}
    \tau_h= \gamma^{-1}  \left(1-\tfrac{\Omega}{a}\right)^{-1}.
\end{eqnarray}
see \eqref{Ch-decay-rate} in the Appendix.
The dependence on $\gamma$ is well-anticipated from the local relaxation time $\tau_r$ , while the additional factor is deduced in Appendix C. The meaning of this factor is that for $a\to\Omega$ when the system is almost unbound, its relaxation time is naturally very large.

\eqref{tau-h} implies that, over timescales much larger than $\tau_h$, white Gaussian noise with intensity $q_h=\int_{-\infty}^{\infty} C_h(t)\d t=\frac{T^2 (a^2+\gamma^2)}{\gamma  (a^2-\Omega^2 ) (a^4+\gamma^2\Omega^2)}$ (c.f. \eqref{Ch-calc})
is a suitable approximation of process $h(t)$.
With this in mind, using \eqref{dL}, we can now formulate the Langevin equation for $L_z \equiv L$:
\begin{eqnarray}
        \tfrac{\d}{\d t}L &=&  
        \la \tfrac{\d}{\d t}L \ra
        +
        4 a^2 \epsilon \,\sqrt{N} \,\sqrt{q_h}\;\xi_h(t)
        \\
        \label{langevin-long}
        &=&\!\!-\!
     \frac{4a^2 \epsilon^2}{\gamma} \,\frac{2\Omega N T}{a^2 \!-\! \Omega ^2}\;\frac{a^2(a^2\!+\!\gamma^2)}{a^4\!+\!\gamma^2\Omega^2} 
     \!+\! 4 a^2 \epsilon\sqrt{\frac{N T^2 (a^2+\gamma^2)}{\!\gamma  (a^2\!-\!\Omega^2 ) (a^4\!+\!\gamma^2\Omega^2)}}\;\xi_h(t),
\end{eqnarray}
This is a stochastic differential equation for $L$, but the right-hand side depends on $\Omega$ and $T$. In the next section, we derive the angular velocity and temperature as a function of the angular momentum: $\Omega(L)$ and $T(L)$, thus \eqref{langevin-long} will be a closed Langevin equation for $L$. 

We make two remarks about the diffusion term in \eqref{langevin-long}. (i) As $\xi_h(t)$ and $-\xi_h(t)$ are statistically the same process, the \eqref{langevin-long} is invariant of the sign of $\epsilon$, i.e. the solution of \eqref{langevin-long} does not depend on the sign of $\epsilon$.
(ii) The diffusion term depends on $L$ (as shown in the next section) and the noise becomes multiplicative, therefore, it is crucial to specify the interpretation of the noise (e.g., It\^{o} or Stratonovich; Ref. \cite{risken} discusses the difference).  Since we obtained \eqref{langevin-long} by approximating a physical colored noise with a white noise, the interpretation of \eqref{langevin-long} should be done via the Stratonovich scheme. (In some cases, it should be possible to directly use colored noise $h(t)$ in Langevin equation \eqref{langevin-long}.) We shall not dwell on proving this point, because the choice of interpretation is not important for $N\gg 1$: the change in the drift term of the \FP{} equation that corresponds to \eqref{langevin-long} is of the order ${\cal O}(N^{-1})$, i.e. negligible for a large number of particles.
Hence, we can also continue with the It\^{o} interpretation. (iii) If we rescale time $t'=t/\tau_L$ with $\tau_L =\frac{\gamma}{4a^2\epsilon^2}$, the Langevin equation reads
\begin{eqnarray}
        \frac{\d}{\d t'}L =\,&-\frac{2\Omega N T}{a^2 - \Omega ^2}\;\frac{a^2(a^2+\gamma^2)}{a^4+\gamma^2\Omega^2} 
     + \sqrt{\frac{4 N T^2}{(a^2-\Omega^2 )}\frac{a^2 (a^2+\gamma^2)}{   (a^4+\gamma^2\Omega^2)}}\;\xi_h(t'). 
     \label{krom}
\end{eqnarray}
Note that for $a\gg \gamma$ and $a^2\gg\gamma\Omega$, (\ref{krom}) will depend only on the temperature and basic dynamical parameters of the system, and not on the small symmetry-breaking parameter $\epsilon$ and/or phenomenological factor $\gamma$ that are normally difficult to control in practice. This is an advantage. We emphasize, however, that (\ref{langevin-long}) is not yet a closed equation. This closing is achieved in the next section.

\section{Canonical and microcanonical distributions}
\label{sec-stationary}

We start this section by showing how statistical ensembles in (\ref{microcanonical}) and (\ref{canonic}) (as well as their parameters) are related to each other \cite{terletski,berdichevsky}. This will allow us to connect the microcanonical quantity $L$ to the canonical parameters $T$ and $\Omega$, and hence to establish \eqref{langevin-long} as a closed stochastic differential equation for $L$.

Let us recall that the overall closed $N$-particle system is partitioned into a subsystem and a bath. For the Hamiltonian this implies
\begin{eqnarray}\label{H-partition}
    H(\psv_\text{sub}, \psv_\text{bath})=H_\text{sub}(\psv_\text{sub})+H_{\rm bath}(\psv_\text{bath})
    +H_\text{interaction}(\psv_\text{sub}, \psv_\text{bath}),
\end{eqnarray}
where $\psv_\text{sub}$ and $\psv_\text{bath}$ are canonical variables (coordinates and momenta) of the subsystem and bath, respectively. Together they unambiguously define the state of the whole system.
Consider the stationary microcanonical ensemble \eqref{microcanonical} for a large system. (Recall that it is stationary due to Liouville's theorem \cite{landau1960mechanics}.)
Marginalization of this distribution over variables of the bath results in a distribution function for the subsystem. Under certain conditions \cite{terletski, berdichevsky, touchette}, the marginalized distribution is the Gibbs canonical distribution \eqref{canonic}. In other words, for any function $f(\psv_\text{sub})$ we have 
\begin{equation}\label{ensamble-equiv}
\la f(\psv_\text{sub})\ra_\text{m}=\la f(\psv_\text{sub})\ra_\text{c},    
\end{equation}
where 
$\la ~\ra_\text{m}$ and $\la ~\ra_\text{c}$ denote averaging w.r.t. microcanonical \eqref{microcanonical} and canonical \eqref{canonic} distribution, respectively.

Equation (\ref{ensamble-equiv}) holds under the following conditions.
(i) A large number of degrees of freedom: $N\gg 1$. The mechanism behind this assumption is the concentration of the volume of phase space near the surface $H=E$. Thus, the entropy, which is defined by the phase-space volume $H\le E$ can be approximated by the area of $H=E$ surface. (ii)  A small subsystem such that $H_\text{sub}(\psv_\text{sub})/E\ll 1$. This ratio is used as a small parameter and the canonical distribution is the first-order expansion in terms of that small parameter. (iii) A weak interaction between the subsystem and the rest of the system. This condition does not mean the complete exclusion of interactions, as it is the only mechanism for establishing thermodynamic equilibrium \footnote{There are systems where thermalization happens due to mixing and collisionless relaxation. In \ref{app-collisionless} we show that thermalization does not happen in a collisionless system.}.
We just require $H_\text{interaction}$ to be negligible compared to other terms in \eqref{H-partition}.
This is necessary to write the energy of the bath only by the Hamiltonian of the subsystem.
This condition is also called  ``additivity of energy''. (iv) Recall that, on top of this, we also need the system to be ergodic to be able to discuss thermodynamics, including concepts like temperature or relaxation.

Essentially, the equivalence is only established in the asymptotic limit $N\rightarrow\infty$ and $H_\text{sub}/E\rightarrow 0$. If the interactions are short-range, the third condition will be automatically fulfilled if the subsystem also has many degrees of freedom.

In our case, the total energy and the $z$ component of the total angular momentum of the system reads:
\numparts
\begin{eqnarray}\label{E-def}
    E &=& U_\text{int}(\{\r_i\})+\sum_{i=1}^{N} \left(
    \tfrac{1}{2}\v_i^2+V(\r_i)
    \right),
    \\ 
    \label{L-def}
    L &=& \sum_i (\r_i \times \v_i)_z=\sum_i \left( 
    x_iv_{y\,i}-y_i v_{x\,i}
    \right).
\end{eqnarray}
\endnumparts
Here, $\r=(x,y,z)$ represents the position and $\v=(v_x,v_y,v_z)$ represents the velocity of the particle; $(\r\times\v)_z$ denotes $z$ component of the cross-product between them. The function  $U_\text{int}(\{\r_i\})=\sum_{i<j}{\cal U}\left(\mid\r_i-\r_j\mid\right)$ is the interaction between particles, $V(\r)$ is the external confining potential and $N$ is the number of particles.
We consider the quadratic potential \eqref{potential-perturbed}, but neglect the effects of anisotropy:
\begin{eqnarray}\label{potential-sym}
    V(\r)=&\frac{1}{2}a^2 x^2+\frac{1}{2}a^2 y^2+\frac{1}{2}a_z^2 z^2,
\end{eqnarray}
so this has rotational symmetry exclusively along the $z$-axis so that only the $L_z$ is conserves. The perturbed potential \eqref{potential-perturbed} introduces corrections of order $\epsilon$ which are negligible in this analysis.
Let the subsystem be just one of the particles.

The left-hand side of \eqref{L-def} is conserved, and if we take the ensemble average of both sides, we directly get $L = \la L \ra_\text{m}=N \la L_1 \ra_\text{m} = N \la xv_y-yv_x\ra_\text{m}=N \la xv_y-yv_x\ra_\text{c} $.
Here, again, $\la ~\ra_\text{m}$ and $\la ~\ra_\text{c}$ denote averaging w.r.t. microcanonical \eqref{microcanonical} and canonical \eqref{canonic} distribution, respectively.
Additionally, we used the ensemble equivalence and the fact that the average of the sum is the sum of averages.
For energy equation \eqref{E-def} we need the assumption of the additivity of energy discussed at the beginning of the section. Rewrite \eqref{E-def} as
\begin{eqnarray}\label{EL-def-sub}
    &E =U_\text{int}(\{\r_i\})+\sum_{i=1}^{N}E_1(\r_i,\v_i),
    \\
    &E_1(\r,\v)=\tfrac{1}{2}\v^2+V(\r),
\end{eqnarray}
where $E_1$ is the energy of the subsystem consisting of one particle. Now, we take the ensemble average of both sides 
\begin{eqnarray}{2}
    E&=&\la E\ra_\text{m}
    \\
    &\approxeq &\,N\la E_1 \ra_\text{m} \qquad \text{(energy\;additivity;\,neglected\;weak \;interactions)}
    \\
    &=&\,N\la E_1 \ra_\text{c} \qquad \text{(equivalence\;of\;ensembles)}
    \label{ormuzd}
\end{eqnarray}
For one-particle subsystem, the canonical distribution \eqref{canonic} reduces to
\begin{eqnarray}\label{canonic-2}
    \rho_\text{c}(\r,\v)\propto \exp\left[-\tfrac{1}{T}\left( \tfrac{1}{2}\v^2+V(\r)-\Omega (xv_y-yv_x) \right)\right].
\end{eqnarray}
Performing the average with (\ref{canonic-2}) in (\ref{ormuzd}), we get
\begin{eqnarray}\label{EL-from-TO}
    E(T,\Omega) = NT \, \frac{3a^2-\Omega^2}{a^2-\Omega^2}
    \quad\text{and}\quad
    L(T,\Omega)=NT\frac{2\Omega }{a^2 - \Omega ^2},
\end{eqnarray}
and solving these for for $\Omega$ and $T$:
\begin{eqnarray}
\comment{T(E, \Omega) = \frac{E}{N} 
    \label{TL-from-EO}
    \frac{a^2-\Omega ^2}{3 a^2-\Omega ^2},
        \qquad
        L(E, \Omega) = E\frac{2\Omega }{3 a^2 - \Omega ^2},
    \\} 
    \label{TO-from-EL}
    \Omega(L, E)=\frac{E}{L}\left(\sqrt{1+\frac{3 a^2 L^2}{E^2}}-1\right),
    \\
    T(L, E)=\frac{E}{3N}\left(2 - \sqrt{1+\frac{3 a^2 L^2}{E^2}}\right),
\end{eqnarray}
where we directly see that during the decay of angular momentum, the temperature $T$ increases. In other words, the ordered rotational motion is converted into the chaotic motion of particles. Similar identities for the general potential $V(\r)$ are derived in Ref. \cite{laliena}.

There is a divergence and unphysical behavior in the aforementioned expressions when $\Omega \rightarrow a$. 
To understand this, consider the marginal distribution for the position $\r$: from \eqref{canonic-2} we have 
\begin{eqnarray}
    \rho_\text{c}(\r)=\int \d \v\, \rho_\text{c}(\r,\v)
    &\propto& \exp\left[-\frac{1}{T}\left(V(\r)-\frac{\Omega^2}{2}(x^2+y^2)\right)\right]
    \\
    &=&\exp\left[-\frac{a^2-\Omega^2}{2T}(x^2+y^2)-\frac{a_z^2}{2T}\,z^2\right]
    \label{rho-x}
\end{eqnarray}
When $\Omega \ge a$ the above distribution is not well defined (non-vanishing at $x^2+y^2\rightarrow\infty$).
This is due to the centrifugal force in the rotating frame, which creates additional potential $-\frac{1}{2}\Omega^2 (x^2+y^2)$, see \ref{app-derivation}. When $\Omega \ge a$ the total potential ceases to be confining, and there is no steady state.

Equation \eqref{rho-x} answers another interesting question. The system can expand either due to rotation (centrifugal forces) or by the increase in the temperature. During the relaxation, the rotation slows down, and the system heats up.  Does the system contract or expand during the relaxation? From \eqref{rho-x} we have
\begin{equation}
    \la x^2 +y^2 \ra = \frac{T}{a^2-\Omega^2} = \frac{E/N}{3a^2-\Omega^2},  \qquad \la z^2 \ra = T = \frac{E}{N}\frac{a^2-\Omega^2}{3a^2-\Omega^2}
\end{equation}
This shows that the system monotonically contracts in $x$-$y$ plane, but expands along $z$ direction.

\section{Numerical Simulations}
\label{sec-numerics}

\begin{figure}[t]
\centering
\begin{minipage}[t] {.5\textwidth}
  \centering
  \includegraphics[width=\linewidth]{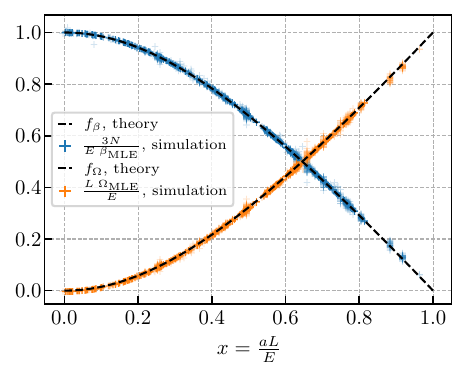}
  \caption{
  Verification of the relationships between parameters of microcanonical and canonical distributions. 
  The data points are computed from independent simulations, while the dashed lines correspond to functions defined in \eqref{fdef}. This analysis demonstrates that we have ensemble equivalence \eqref{ensamble-equiv} in the case of low density.
  }
  \label{fig-compare-1}
\end{minipage}%
\hfill
\begin{minipage}[t]{.45\textwidth}
  \centering
  \includegraphics[width=0.9\linewidth]{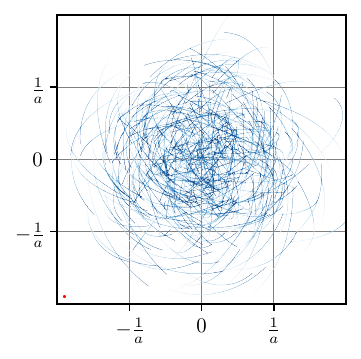}
  \caption{
Trajectories of 523 particles over 5 units of time projected on $xy$ plane, corresponding to the system of Figure \ref{fig-L} at $L\simeq1500$. The gradient of the trajectory indicates the direction of time. The system is in the underdamped regime. 
  The small red dot (at the bottom left corner) indicates the size of the Lennard-Jones particle.}
  \label{fig-snap}
\end{minipage}
\end{figure}

In this section, we present the results of our numerical molecular dynamic simulations, which are conducted to validate the mesoscopic descriptions provided in equations \eqref{langevin} and \eqref{FP0}, as well as the coarse-grained stochastic differential equation (SDE) in \eqref{langevin-long}. The simulations focus on $N=523$ particles confined in an external quadratic potential \eqref{potential-perturbed}. Additionally, to ensure system relaxation and thermalization, we introduce weak particle-particle interactions governed by the Lennard-Jones potential:
\begin{equation}\label{lennard}
    {\cal U}(r) = 4\varepsilon \left[ -(\sigma_0/r)^6+(\sigma_0/r)^{12}\right],
\end{equation}
where, $\sigma_0$ is referred to as the diameter of the particle. More precisely, this is the size of the soft repulsive core.
The time average of the interaction energy is 0.1\% of total energy so that $H_\text{interaction}$ is negligible in \eqref{H-partition}.

The details about figures and simulation parameters can be found in \ref{app-numerics}.

\subsection{Verification of the ensemble equivalence}
To investigate the relationships presented in equations \eqref{EL-from-TO} and \eqref{TO-from-EL}, we express them in dimensionless form:
\begin{eqnarray}
\label{fs}
    \frac{3N\,T}{E}=f_\beta\left(\frac{a L}{E}\right),
    \qquad
    \frac{\Omega L}{E}=f_\Omega\left(\frac{a L}{E}\right),
    \\
    \label{fdef}
    f_\beta(x)=\frac{3(1-x^2)}{2+\sqrt{1+3x^2}},
    \quad
    f_\Omega(x)=\sqrt{1+3x^2}-1.
\end{eqnarray}
We perform simulations by varying the potential parameter $a$, the particle number $N$, and initial values for the conserved quantities $E$ and $L$. For each simulation, we estimate the temperature and angular velocity using the Maximum Likelihood Estimators defined in \ref{app-TO}. Using these estimates, we calculate the argument $x=\frac{aL}{E}$ and the left-hand sides of \eqref{fs}, which we then compare with the expected functional relationships in \eqref{fs}.
Figure \ref{fig-compare-1} illustrates the consistency between our numerical results and the analytical predictions when the system density is low, $\rho \ll \sigma_0^{-3}$, where $\sigma_0$ is the characteristic size of the particles (also see \eqref{lennard}).
The high-density case is discussed in the following section.

\subsection{Simulations with high density} \label{sec-high-dens}
We repeat the analysis of the previous section, but now we take a 1000 times denser system and $\rho \lessapprox \sigma_0^{-3}$, i.e. the particles are 10 times closer to each other on average. This corresponds to the liquid phase of the Lennard-Jones model. The result is shown in Figure \ref{fig-compare-dense} which indicates the complete failure of relations in \eqref{fs} in high-density regime.

Apparently, in this case, the microcanonical and canonical distributions are not equivalent.
The canonical distribution \eqref{canonic-2} is Gaussian \emph{only} when the potential is quadratic (see \eqref{potential-sym}). 
We test the distributions of $\r$ and $\v$ in the dense system
by quantile-quantile plots (Q-Q plot \cite{thode2002testing}): we standardized (subtracted the mean and scaled to unit variance) velocities and positions, then compared them to the standard normal distribution. 
The outcome, illustrated in Figure \ref{fig-QQ}, clearly demonstrates that neither $v_y$ nor $x$ adheres to a Gaussian distribution, while the results in \eqref{EL-from-TO} require Gaussian distribution. 
However, $v_z$ is still Gaussian because it is not coupled to the position $\r$ in canonical distribution \eqref{canonic-2}. This means that for $L=\Omega=0$ (i.e. when the coupling in \eqref{canonic-2} vanishes) we would expect Gaussian distributions for all velocities. 

The finite size of the particles sets an upper bound on the density. Close to this bound, particles are closely packed and one of the crucial assumptions for the equivalence of microcanonical and canonical distributions is violated. Particle-particle interactions create an effective potential that a single particle feels in a steady state, see the $H_\text{interaction}$ term in \eqref{H-partition}.
As a workaround, the external potential in \eqref{canonic-2} can be replaced by an effective potential. 
For example, in the derivation of the Generalized Langevin, the potential mean force (PMF) replaces the external potential \cite{langevingeneralized}. However, Figure \ref{fig-QQ} indicates non-gaussian distribution for $\r$, resulting in anharmonic potential mean force,  whose analytical form is unknown.

We continue our simulations in the low-density limit as in Figure \ref{fig-compare-1} when \eqref{EL-from-TO} and \eqref{TO-from-EL} are valid.
As a consequence, the friction constant $\gamma$ is small, and the system is in the underdamped regime. In terms of model parameters, $\gamma\ll a$, which means the system makes multiple oscillations before relaxation.

\subsection{Anisotropic Potential: Relaxation of Mean Angular Momentum}\label{sec-mean-relax}

\begin{figure*}[t]
\centering
  \includegraphics[width=\linewidth]{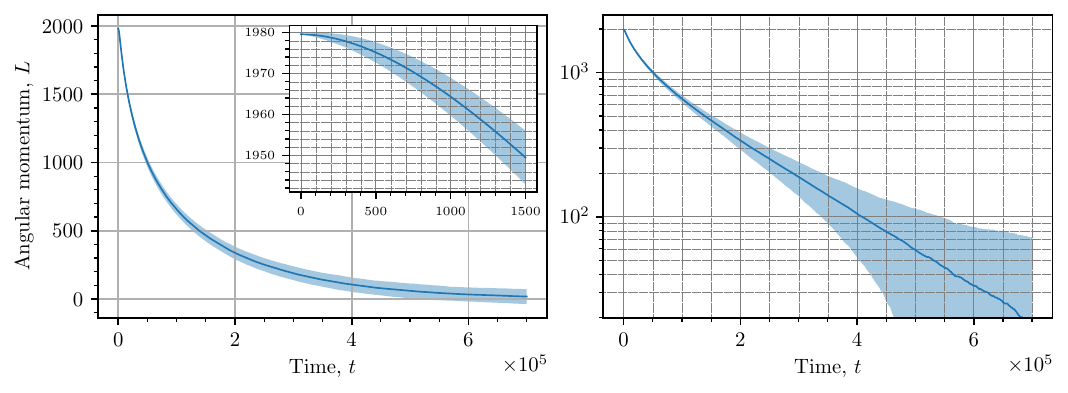}
  \caption{
    This figure illustrates the relaxation of total angular momentum $L$. The solid line is the ensemble average over 1200 simulations with the same initial $E$ and $L$. 
    The shaded region surrounding the line corresponds to one standard deviation of the values within the ensemble.
    The inset shows the relaxation just after the violation of the rotational symmetry. For a different perspective, the same plot is presented on the right side with a logarithmic scale.}
  \label{fig-L}
\end{figure*}

Now we turn to the case when the external potential in (\ref{potential-perturbed}) is anisotropic due to a small $\epsilon$ violating the rotation symmetry along the $z$ axis.

The dynamics of the total angular momentum become stochastic in this case, and we conduct an ensemble of simulations with identical initial values of $E$ and $L$ (see \ref{app-random-EL} for details). The relaxation of $L$ is visualized in Figure \ref{fig-L}. 
The intermediate state of the system is visualized in Figure \ref{fig-snap}.

\begin{figure}
\centering
\begin{minipage}[t]{.49\textwidth}
  \centering
  \includegraphics[width=\linewidth]{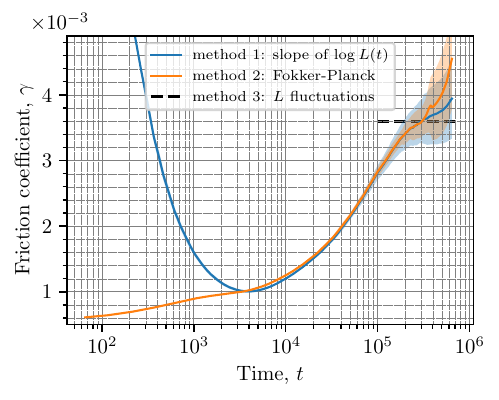}
  \caption{
  Three methods of estimating $\gamma(t)$. The 1st and 2nd methods utilize the dissipation of $L$, while the 3rd method uses fluctuations of angular momentum in steady state to estimate $\gamma$.}
  \label{fig-gamma}
\end{minipage}%
\hfill
\begin{minipage}[t]{.49\textwidth}
  \centering
  \includegraphics[width=\linewidth]{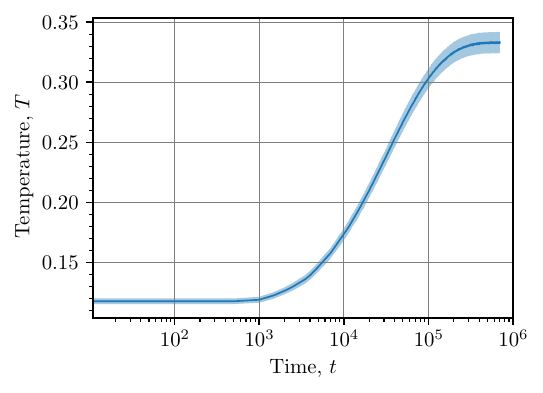}
  \caption{
  Temperature of the system during the relaxation, computed by maximum likelihood estimator. Again, the shaded region is the one standard deviation of estimated values.  Note that temperature increases more than twice.}
  \label{fig-T}
\end{minipage}
\end{figure}

As the friction constant is small ($\gamma \ll a$) in the dilute system, the deterministic part (drift term) of the Langevin \eqref{langevin-long} has a simpler form after the substitution of \eqref{TO-from-EL}:
\begin{eqnarray}\label{L-deterministic}
    \frac{\d}{\d t}\la L \ra = \frac{4a^2 \epsilon^2}{\gamma} \la L \ra 
\end{eqnarray}
The logarithmic plot on Figure \ref{fig-L} suggests that $\gamma$ increases over time. The change of $\gamma$ is acceptable as it can depend on the temperature and density of the system, and both of them change significantly as the angular momentum relaxes to zero. The temperature dependence of the system is presented in Figure \ref{fig-T}. It is apparent from the visualization of the trajectories (Figure \ref{fig-snap}) that system density is not uniform either.

We estimated $\gamma$ using three methods. The simplest one involves calculating the slope of $\log\la L(t)\ra$: we take a small time interval of the order of the local relaxation time $\tau_r$. We choose $\epsilon$ small enough so that $L$ changes by less than $1\%$ within this interval. Then the slope is estimated in this interval and  $\gamma$ is computed from it according to \eqref{L-deterministic}.

The second method does not assume a quasi-stationary state or a small $\gamma$, and is more computationally intensive. 
In this approach, we keep the time derivative in \FP{} \eqref{FP0}.
However, it is important to note that the standard form of \FP{} is inherently designed to describe a subsystem and cannot directly represent an isolated system with conserved energy. This is primarily due to the presence of a diffusion term, which ``injects'' energy into the system with a temperature-dependent rate: if the temperature is high, energy will be injected, while at low temperatures, dissipation dominates and the total energy will decrease.
We modify the \FP{} by introducing temperature as a dynamical variable: at each time, the temperature is chosen, such that the average total energy is conserved. The details are presented in \ref{app-gammafit}. Then we fit the $\gamma(t)$ function, such that the solution of the modified \FP{} reproduce observed $L(t)$ in Figure \ref{fig-L}.

The third method uses the fluctuations of $L$ in the steady state and is discussed in the following section. This estimate does not have temporal behavior like the other methods.
 
The results of these three methods are presented in Figure \ref{fig-gamma}.
Notably, the first method exhibits a divergence at small times, consistent with the inset of Figure \ref{fig-L}: initially, the dynamics follows a ballistic regime with $L(t)\propto \cos(2 a \epsilon t)$ as can be seen by a simple calculation for non-interacting particles; see \ref{app-collisionless}. 
Therefore, the slope of $L(t)$ approaches zero at the beginning (see inset in Figure \ref{fig-L}), resulting in an infinite estimate for $\gamma$ near time 0. 
There is no equilibration (not even a relaxation) in this collisionless system, even though the total angular momentum decays initially; see \ref{app-collisionless}.
 
After a time of the order of $\tau_r \approx \gamma^{-1}$, the estimates become consistent with the \FP{} method. As the simulation progresses, $\gamma$ estimates become noisier because the standard deviation of $L$ increases over time and the estimated average angular momentum becomes noisier; see Figure \ref{fig-L} at large times. Eventually, the first two estimators approach the third estimator as the system approaches the steady state.

\begin{figure}
    \centering
  \includegraphics[width=0.6\linewidth]{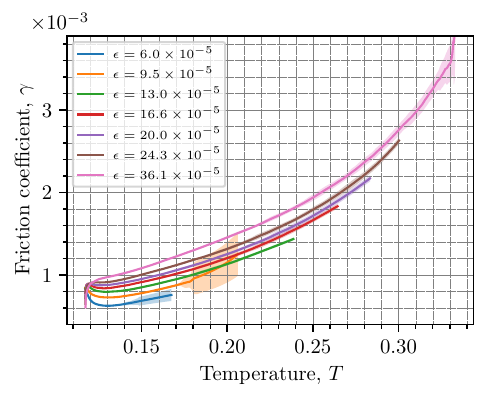}
  \caption{Parametric plot of $\gamma(t)$ and $T(t)$ for varied $\epsilon$ values.
  The simulation interval was the same for all paths and for \textit{small} $\epsilon$ the total $L$ decays \textit{less} during that interval. Thus, temperature variation is also small. This explains the differences in line lengths above.
  }
  \label{fig-gamma-T}
\end{figure}

We anticipate that $\gamma$ depends on various parameters, including the interaction potential, the number of particles, and the density of the system. To systematically explore its temperature dependence in a controlled way, we vary only the degree of anisotropy $\epsilon$ and plot the estimates $\gamma$ from the \FP{} method (2nd method of Figure \ref{fig-gamma}) depending on temperature, i.e. the parametric plot of $T(t)$ and $\gamma(t)$, see Figure \ref{fig-gamma-T}. 
We observe a similar temperature dependence $\gamma(T)$ across different $\epsilon$ values. However, at a fixed temperature, $\gamma$ exhibits a monotonic increase with $\epsilon$, a phenomenon for which we currently lack a conclusive explanation.

It is also worth mentioning that friction can have spatial dependence due to density variations in space (the density is lower at the edges of the system, see Figure \ref{fig-snap}), while we characterize it with a single parameter $\gamma$. In \ref{app-space-gamma}, we consider a space-dependent $\gamma$ that is still tractable and makes predictions.

\subsection{Fluctuations of Angular Momentum}

The Langevin equation \eqref{langevin-long} has multiplicative noise in it. We do not have a method to directly inspect the multiplicative nature of noise in the simulations. 
However, \eqref{langevin-long} predicts the strength of the fluctuations of $L$ in equilibrium. 

The noise in the Langevin equation \eqref{langevin-long} scales as $N^{-1/2}$ as $NT^2$ term in the diffusion term scales as $ N^{-1}$.
Thus, in the thermodynamic limit ($N\gg 1$) the fluctuations of $L$ are small. Therefore, we consider the limit $L\ll E/a$, where
\eqref{TO-from-EL} reads
\begin{eqnarray}\label{OT-approx}
\Omega(L)=\frac{3a^2L}{2E}+{\cal O}(L^2)
\qquad 
T(L)=\frac{E}{3N}+{\cal O}(L^2)
\end{eqnarray}
and the Langevin equation in this limit becomes
\begin{eqnarray}\label{L-langevin-2}
    \tfrac{\d}{\d t}L &=  - \alpha
      L + \sqrt{2 D_L} \;\xi_h(t),
\end{eqnarray}
where 
\begin{eqnarray}\label{alpha-Dl}
    \alpha=\frac{4a^2 \epsilon^2}{\gamma}\frac{a^2+\gamma^2}{a^2} 
    \quad \text{and} \quad 
    D_L=\frac{2E^2}{9Na^2} \frac{4a^2 \epsilon^2}{\gamma}\frac{a^2+\gamma^2}{a^2}
\end{eqnarray}
The above result is for general $\gamma$. In the steady state, $\la L\ra=0$ and 
\begin{eqnarray}\label{L-fluct}
    \la L^2\ra = \frac{D_L}{\alpha} = \frac{2 E^2}{9 a^2 N} 
\end{eqnarray}
Indeed, the fluctuations scale as $N^{-1/2}$.
\eqref{L-fluct} is independent of $\gamma$ and the degree of anisotropy $\epsilon$. 
This result is consistent with equilibrium thermodynamics: in the absence of rotational symmetry, even with weak anisotropy, $L$ is not an integral of motion. The subsystem is no longer described by the canonical distribution in \eqref{canonic}, but rather by $\rho_\text{c}(\r,\v)\propto \exp\left(-\frac{1}{T} E(\r,\v)\right)$.
Certainly, the average of angular momentum $L=xv_y-yv_x$ with respect to this distribution vanishes, and the second moment matches with \eqref{L-fluct} after replacing $T$ with \eqref{OT-approx}.

However, \eqref{L-langevin-2} contains more information than just the second moment. It provides the autocorrelation function of $L(t)$ and the spectrum of the fluctuations:
\begin{eqnarray}
    \la L(t+t_0)L(t_0)\ra = C_L(t)&=&\frac{2 E^2}{9 a^2 N} \;\ee^{-\alpha t},\label{CL-def}
    \\
    \qquad\qquad\quad\qquad S_L(\omega)&=&\frac{2 E^2}{9 a^2 N} \;\frac{2\alpha}{\alpha^2+\omega^2},
    \label{SL-def}
\end{eqnarray}
where $\alpha^{-1}$ is the correlation time of the fluctuations. Due to the stationarity of the equilibrium state, \eqref{CL-def} does not depend on $t_0$. Note that the power spectrum $S_L(\omega)$ is defined as the Fourier Transform of the autocorrelation function $C_L(t)$.

\begin{figure*}[t]
\centering
  \includegraphics[width=\linewidth]{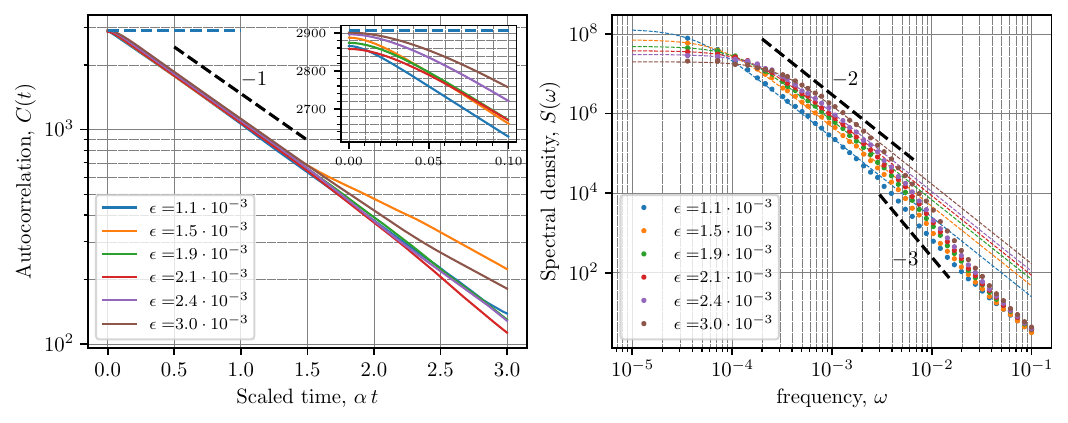}
  \caption{
  \textbf{(left)}
  Autocorrelation function of $L(t)$ in the steady state, when $L$ fluctuates around 0. The horizontal axis corresponds to the time, scaled by decay rate $\alpha$ (c.f. \eqref{alpha-Dl}). The horizontal dashed line corresponds to the steady-state fluctuations and denotes the value of \eqref{L-fluct}. 
  Inset provides a magnified view of the same plot at smaller times.
  \textbf{(right)} Calculated spectral density (points) is compared to the analytical expression (dashed lines), as defined in \eqref{SL-def}.}
  \label{fig-CS}
\end{figure*}

To investigate these fluctuations, we initiate simulations with an initial angular momentum of $L=0$ and allow the system to evolve over a time interval $\tau_L$ until it reaches a steady state. Subsequently, we estimate both the autocorrelation function and the spectral density of the recorded stochastic process $L(t)$.
In Figure \ref{fig-CS}, we present a comparison between the analytical predictions outlined in equations (\ref{CL-def}, \ref{SL-def}) and the estimates derived from our simulations. Notably, we find a strong agreement in the behavior of the autocorrelation function, particularly within the time interval $\tau_r < t < \tau_L$. Any inconsistencies at larger time scales are likely attributable to finite sample size effects, while at smaller times, we observe a ballistic regime reminiscent of the behavior illustrated in Figure \ref{fig-L}.

As for the spectral density $S(\omega)$, we observe a favorable match at low frequencies, specifically when $\omega < \alpha$, corresponding to longer time scales. Within this regime, the white noise approximation of $h(t)$ holds, resulting in a power spectrum with a characteristic $-2$ power-law of the Lorentzian spectral density in \eqref{SL-def}.
However, at higher frequencies, representing shorter time scales, we observe deviations from the white noise approximation, attributed to non-trivial correlations of $h(t)$ coming into play.

The friction coefficient $\gamma$ is estimated from autocorrelation function through the estimation of $\alpha$ in \eqref{alpha-Dl}. The estimated value is shown in Figure \ref{fig-gamma} by the dashed line.

\section{Conclusions}

We presented Langevin and \FP{} equations, which correctly describe a rotating equilibrated system. They depend on the parameters characterizing the canonical ensemble, such as the temperature and rotation speed.
Then, we solved the \FP{} in the case of a perturbed external potential. The solution is for general finite perturbation, however, small perturbations were considered to have timescale separation, so that the solution has a quasi-stationary distribution.
We coarse-grained the angular momentum--toque balance equation and derived the corresponding Langevin equation.
However, the equation depended on the parameters of canonical distribution. We established their dependence on microcanonical parameters -- energy and angular momentum -- and the Langevin equation became a closed stochastic differential equation for angular momentum. 
Hence, the dissipation and fluctuations of the total angular momentum are related to the mesoscopic dissipative coefficient.
Our results were validated by numerical molecular dynamics  simulations, where we considered an ensemble with the same initial energy and total angular momentum. We show that a simple Markovian model (the Langevin equation for the rotating system) describes the macroscopic dynamics reasonably well.

Here are a few remarks that were not initially obvious to us in the beginning.
The system is not just in a steady state characterized by slowly decaying angular momentum. Instead, it is in a quasi-stationary state, which is non-equilibrium in the sense that fluctuations of $L$ grow over time (see Figure \ref{fig-L}) and saturate only when the system equilibrates at $L=0$.
Moreover, the quasi-stationary state differs from the ``frozen'' microcanonical distribution (\eqref{microcanonical} with anisotropic potential). This is because such states do not exhibit decay in angular momentum, as demonstrated in \ref{app-micro-avg}.

Looking ahead, this work lays the foundation for future research. Exploring non-uniform friction constants as a generalization of the model (\ref{app-space-gamma}), considering collisionless relaxation scenarios (\ref{app-collisionless}), and investigating the implications of a trivial microcanonical ensemble (\ref{app-micro-avg}) open avenues for further theoretical developments. These aspects, along with a detailed exploration of the simulation results presented in the appendices, provide opportunities for deeper understanding and refinement of the presented model.

In conclusion, our study contributes to the understanding of 
the rotational states in perturbed symmetries, 
approximate modeling of the system in terms of modified Langevin and \FP{} equations, 
and, as an example, coarse-graining of the stochastic dynamics of the total angular momentum.

Looking ahead, a more practical question is the work extraction from the rotating systems. ATPase is an important example, which extracts the rotational energy in the form of ATP molecules. In our study, all rotational energy transforms into thermal energy. One might ask how effectively one can extract the work out of it. Another pertinent follow-up study is the determination of the correct adiabatic invariant (entropy) for the rotating system. It is also relevant to the study of optimal heat engines.

\ack
I am grateful to Armen Allahverdyan for useful discussions and constant support. This work was supported by SCS of Armenia Grants No. 22AA-1C028 and 21AG-1C038.


\section*{References}

\bibliographystyle{unsrt}
\bibliography{refs}
\appendix
\setcounter{section}{0}

\section{Derivation of mesoscopic phenomenological equations}\label{app-derivation}
\label{app-a}

We get the Langevin equation \eqref{langevin} by replacing the friction force to be proportional to the relative velocity. In this section, we consider another approach, which leads to the same mesoscopic descriptions \eqref{langevin} and \eqref{FP0}.

We shift to the frame of reference rotating with the same angular velocity $\Omega$ as the system, i.e. co-rotates with the system.
Let $\r'=(x',y',z')$ be the particle's coordinates in this frame of reference, and the particle velocity will be $\v'=\frac{\d \r'}{\d t}$. They are related to the position and velocity in the rest frame by the following relations:
\begin{eqnarray}
    \phantom{\Rightarrow}~\r &=& \R_t\r', \label{xtransform}
    \\ 
    \Rightarrow \v &=& \R_t \v' + \o\times \r
    \\
    &=& \R_t \v' + \o\times (\R_t \r'),
    \label{vtransform}
\end{eqnarray}

where $\R_t \equiv \R(\o t)$ and $\R(\bs{\alpha})$ is a rotation matrix which rotates vectors around
axis $\bs{\alpha}$ at angle $|\bs{\alpha}|$.
In this rotating frame, the system is at rest, and we can apply the usual Langevin equation, also accounting for inertial forces. Therefore, the following forces act on the particle in the rotating frame of reference:
\begin{itemize}
    \item $-\nabla_{\r'} V(\r)=-\nabla_{\r'} V(\R_t\r')$ -- deterministic force
depending on the position; this comes from external potential and mean field
created by other particles,
\item $-2 \o \times \v'$ -- Coriolis force, which also depends on relative
velocity,
    \item $-\o \times( \o \times \r' )$ -- the centrifugal force, which acts as
an additional external potential.
    \item $-\gamma  \v'$ -- deterministic friction force, which is defined by
the velocity in the rotating frame,
    \item $\sqrt{2 \gamma T} \;\bxi(t)$ -- stochastic thermal noise term.
\end{itemize}
The noise and friction model the interactions with the neighboring particles, and note that they are related by the fluctuation-dissipation relation. After adding up all the forces, 
the Langevin equation reads:
\begin{eqnarray}\label{langevin-rot}
    \tfrac{\d}{\d t} \r' =& \v',
    \qquad
    \\
    \tfrac{\d}{\d t} \v' =& 
    -\nabla_{\r'} V(\R_t\r')
    -2\, \o \times \v'
    -\o \times( \o \times \r' )
    \\
    &-\gamma \v'
    +\sqrt{2\gamma T} \;\bxi(t) .
\end{eqnarray}
And, with standard approach \cite{risken}, we can write the \FP{} equation:
\begin{eqnarray}\label{FP1}
    \tfrac{\d}{\d t} P' &=&
    -\nabla_{\r'} \left[ \v' P' \right]    
    \\
    \nonumber &&+\nabla_{\v'}  \left[ \left(\nabla_{\r'} V(\R_t \r')
    +\gamma \v'
    +2 \, \o \times \v'
    +\o \times( \o \times \r' )\right) P' \right]
    \\
    &&+ \gamma T \,\nabla_{\v'}^2 P',
\end{eqnarray}
where $P' = P'(t, \r', \v')$ is the probability density function in the rotating frame. It can be shown, e.g. by substitution, that in the case of \emph{rotational
symmetric} $V(\r)$ the stationary distribution is
\begin{eqnarray}\label{stationary-simmetric-rotating}
    P'_{st} &\propto& \exp\left\{-\frac{1}{T}\left( 
    \frac{1}{2} \v'^2 + V(\r') - \frac{1}{2} \Omega^2 (x'^2 + y'^2)
    \right)\right\} 
    \\
    &=&\exp\left\{-\frac{1}{T}\left( 
    {\text{kinetic}\atop \text{energy}}\;+\;{\text{potential} \atop \text{energy}}
    \right)\right\}.
\end{eqnarray}
It is worth mentioning, that this result is consistent with the famous \BvL{} theorem \cite{vanvleck,kaplan2009bohr}: the Coriolis force $-2\,\o \times \v'$ is of the form of the Lorentz force from the magnetic field. According to the theorem, a constant magnetic field should not affect the stationary distribution, which is indeed the case in \eqref{stationary-simmetric-rotating}.

The probability density $P'(t, \r', \v')$ in the rotating frame is related to the probability density $P(t, \r, \v)$ by the following relation:
\begin{equation}\label{prob-transform}
    P'(t, \r', \v') = P\left(t,~ \R_t\r' ,~ \R_t \v' + \o\times (\R_t
\r')~\right),
\end{equation}
where the transformation in (\ref{xtransform}-\ref{vtransform}) is used. Note that the determinant of the transformation is equal to one. If we substitute this into \FP{} (\ref{FP1}),
we get a partial differntial equation for $P(t, \r, \v)$:
\begin{eqnarray} \label{FP0-app}
    \tfrac{\d}{\d t} P&=& 
    -\nabla_{\r} \left[ \v P \right]
    +\nabla_{\v}  \left[ \left(\nabla_{\r} V(\r)
    +\gamma (\v-\o\times\r) \right) P \right]
    + \gamma T \,\nabla_{\v}^2 P,
\end{eqnarray}
which is exactly \eqref{FP0}.
The equivalent Langevin equation is \eqref{langevin}. 

Note that \eqref{FP0-app} is different from the usual Fokker-Plank equation for the
non-rotating system by the drift term: now, the friction is proportional to the relative velocity $(\v-\o\times\r)$. It is noteworthy that even though in the rotating frame there is velocity-dependent Coriolis force, and we start with the Langevin equation in that frame, only the drift term is altered in the final Langevin equation \eqref{langevin}.

The following identities were used to transform \eqref{FP1} to \eqref{FP0-app}:
\begin{eqnarray}
    \nabla_{\v'}P' &=& \R_{-t} \nabla_{\v} P,
    \\
    \nabla_{\r'} P' &=& \R_{-t}\nabla_{\r}P-\R_{-t} (\o\times\nabla_{\v}P),
    \\
    \partial_t P'&=&\partial_t P + (\o\times\R_{t}\r')\cdot\nabla_{\r}P+(\o\times(\R_{t}\v'+\o\times\R_{t}\r'))\cdot\nabla_{\v}P \nonumber
    \\
    &=&\partial_t P + (\o\times\r)\cdot\nabla_{\r}P
    +(\o\times\v)\cdot\nabla_{\v}P.
\end{eqnarray}
Note that $\R_{t}^\intercal=\R_{-t}$, 
$\R_{-t}\R_{t}=\bs{I}$ and $\R_{t}(\o\times\bs{a})=\o\times\R_{t}\bs{a}$
for any vector $\bs{a}$.

\section{Solution of the \FP{} of Ornstein-Uhlenbeck Process}\label{app-FP-OU}

We consider the following \FP{} equation (summation with repeated indices):
\begin{equation}
\label{FP-OU}
\frac{\partial P}{\partial t}=\Upsilon_{i j} \frac{\partial}{\partial \rx_i}\left(\rx_j P\right)+ {\rm D}_{i j} \frac{\partial^2 P}{\partial \rx_i \partial \rx_j},
\end{equation}
where $P=P(t, \brx)$ is a distribution function over $\brx=(\rx_1,\dots,\rx_N)$ variables at time $t$.
Here $\Upsilon$ and $\rD$ are matrices of constant coefficients. Moreover, $\rD$ has to be positive semi-definite.
This corresponds to the Ornstein-Uhlenbeck Process 
\begin{eqnarray}
&\dot{X_i}+ \Upsilon_{i j} X_j=\Gamma_i(t) ; \quad i=1, \ldots, N,
\\
&\left\langle\Gamma_i(t)\right\rangle=0, \quad\left\langle\Gamma_i(t) \Gamma_j\left(t^{\prime}\right)\right\rangle=2 \rD_{i j} \delta\left(t-t^{\prime}\right), \quad \rD_{i j}=\rD_{j i},
\end{eqnarray}
where $\left(\Gamma_i\right)_{i=1}^{N}$ is a white Gaussian noise with covariance matrix $2\rD$.
We start with the following initial distribution:
\begin{equation}
P\left(\brx, 0 \mid \brx^{\prime}\right)=\delta\left(\brx-\brx'\right).
\end{equation}
If we express $P$ by its Fourier transform with respect to $\brx$, i.e., by
\begin{equation}
P\left(\brx, t \mid \brx'\right)=\frac{1}{(2 \pi)^{N}} \int \ee^{\ii \left(k_1 \rx_1+\ldots+k_N \rx_N\right)} \tilde{P}\left(\k, t \mid \brx'\right) \d \k.
\end{equation}
In terms of $\tilde{P}$ the \FP{} \eqref{FP-OU} reads
\begin{equation}
    \frac{\partial \tilde{P}}{\partial t}=-\Upsilon_{i j}  k_i
    \frac{\partial \tilde{P}}{\partial k_j}
    -\rD_{i j} k_i k_j \tilde{P} 
\end{equation}
with initial condition
\begin{equation}
    \tilde{P}\left({k}, 0 \mid \brx^{\prime}\right)=\ee^{-\ii \k\cdot \brx'}.
\end{equation}
The solution of this equation is \cite{risken}:
\begin{equation}\label{solution-fourier}
    \tilde{P}\left(\k, t \mid \brx^{\prime} \right)=\exp \left[-\ii k_i M_i\left(t\right)-\frac{1}{2} k_i k_j \sigma_{i j}\left(t\right)\right],
\end{equation}
where $M_i$ and $\sigma_{ij}$ satisfy
\begin{eqnarray}
\dot{M}_i&=&-\Upsilon_{i j} M_j,
 M_i(0)=\rx'_i ,
\\
\dot{\sigma}_{i j}&=&-\Upsilon_{i l} \sigma_{l j}-\Upsilon_{j l} \sigma_{l i}+2 \rD_{i j},
\qquad
\sigma_{ij}(0)=0.
\end{eqnarray}
Their closed-form solutions are
\begin{eqnarray}
M_i(t)&=G_{i j}(t) \rx_j^{\prime}, 
\quad
\sigma_{i j}(\tau)=\int_0^\tau 
2 \rD_{k l} \,{\rm G}_{i k}(s) {\rm G}_{j l}\left(s\right) \mathrm{d} s ,
\label{sdigma-def}
\end{eqnarray}
where ${\rm G}_{i j}(t)$ is the Green's function and reads (matrix form)
\begin{equation}
    {\rm G}(t)=\exp (-\Upsilon t)=\brI-\Upsilon t+\frac{1}{2} \Upsilon^2 t^2 + \ldots\;,
\end{equation}
where $\brI$ is the identity matrix.
Solution \eqref{solution-fourier} in real space will be the propagator, ${P}\left(\brx, t \mid \brx^{\prime} \right)=\mathcal{N}(\brx;\;M(t),\sigma(t))$, where $\sigma(t)$ is the covariance matrix from \eqref{sdigma-def}.
If at time $t=0$, the distribution is Gaussian with mean $\bs\mu_0$ and variance $\Sigma_0$:
\begin{equation}
    P(\brx,0)=\mathcal{N}(\brx;\;\bs\mu_0,\Sigma_0),
\end{equation}
then, using the propagator, it can be shown that
\begin{eqnarray}
    P(\brx,t)&=&\int P(\brx,t\mid \brx')P(\brx',0)\d \brx'
    \\
    &=&\mathcal{N}\left(\brx;\;\bs\mu={\rm G}(t)\bs\mu_0,\;\Sigma={\rm G}(t)\Sigma_0 {\rm G}(t)^\TT+\sigma(t)\;\right).
    \label{FP-full-sol}
\end{eqnarray}
The steady-state solution will be $P_{st}(\brx)=P(\brx,t\rightarrow\infty|\brx')=P(\brx,t\rightarrow\infty)$
\begin{eqnarray}
    P_{st}(\brx)= \mathcal{N}\left(\brx;\;\bs\mu=0,\;\Sigma=\sigma_\infty\;\right),\label{FP-OU-full-sol}
\end{eqnarray}
where $\sigma_\infty$ solves matrix equation 
\begin{eqnarray}\label{stationary-cond}
    \Upsilon\sigma_\infty+\sigma_\infty\Upsilon^\TT=2 \rD.
\end{eqnarray} 

\subsection{(Quasi-)Stationary solution for our system}
From \FP{} \eqref{FP0}, with $\brx=(\r, \v)$ being six-dimensional vector, we read off the $\Upsilon$ and $\rD$ coefficients

\begin{eqnarray}\label{FP-coefs}
\Upsilon = \pmatrix{
    \gamma\; \brI & {\rm A}-\gamma\uo
    \cr
    -\brI & 0},
\qquad
\rD=\pmatrix{
    \gamma T\,\brI & 0
    \cr
    0 & 0},
\\
\text{where}\quad 
{\rm A}=\pmatrix{
    a^2(1\!+\!\epsilon)^2 \!\!\!\!\!&0&0\cr
    0&a^2(1\!+\!\epsilon)^2\!\!\!\!
    &0\cr
    0&0&a_z^2}
\qquad
\uo=\pmatrix{
    0&-\Omega&0\cr
    \Omega&0&0\cr
    0&0&0}
\end{eqnarray}
and $\brI$ is $3\times 3$ identity matrix. Solving \eqref{stationary-cond} with the above coefficients we get
\begin{eqnarray}
    \sigma_\infty = \pmatrix{
        \Sigma_{\r\r} & \Sigma_{\r\v}
        \cr
        \Sigma_{\r\v}^\TT & \Sigma_{\v\v}},
\end{eqnarray}
\begin{eqnarray}\label{sigma-rr}
    \Sigma_{\r\r} =& T 
\pmatrix{
\frac{1}{a^2-\Omega^2}-\frac{2\epsilon a^2}{a^4+\gamma^2 \Omega^2}
& -\frac{2\epsilon a^2 \Omega \left(a^2+\gamma^2\right) }{\gamma(a^2-\Omega^2)\left(a^4+\gamma^2
\Omega^2\right)} 
& 0
\cr
-\frac{2\epsilon a^2 \Omega \left(a^2+\gamma^2\right) }{\gamma(a^2-\Omega^2)\left(a^4+\gamma^2
\Omega^2\right)} 
& \frac{1}{a^2-\Omega^2}+\frac{2\epsilon a^2}{a^4+\gamma^2 \Omega^2}
& 0
\cr
0 & 0 & \frac{1}{a_z^2}
}
+ O\left(\epsilon^2\right),
\end{eqnarray}
\begin{eqnarray}\label{sigma-vv}
\Sigma_{\v\v} = T
\pmatrix{
\frac{ a^2}{a^2-\Omega^2} 
& -\frac{ 2\epsilon \Omega a^2}{\gamma(a^2-\Omega^2)}
& 0
\cr
-\frac{ 2\epsilon \Omega a^2}{\gamma(a^2-\Omega^2)}
& \frac{ a^2}{a^2-\Omega^2}
& 0
\cr
0 & 0 & 1
}
+ O\left(\epsilon^2\right),
\end{eqnarray}
\begin{eqnarray}\label{sigma-rv}
\Sigma_{\r\v} = \frac{T\;\Omega}{a^2-\Omega^2}\pmatrix{
    0&1&0\cr 
    -1&0&0\cr
    0&0&0
    }+ O\left(\epsilon^2\right) .
\end{eqnarray}
In our analysis we only need one of the components:
\begin{eqnarray}
    \Sigma_{xy}=\la x y\ra &=& -\frac{2\epsilon T a^2 \Omega \left(a^2+\gamma^2\right) }{\gamma(a^2-\Omega^2)\left(a^4+\gamma^2
\Omega^2\right)} + {\cal O}(\epsilon^2)
\\
\label{xy-mean-app}
    &=& -L \, \frac{a^2 \epsilon}{N \gamma} \, \frac{a^2+\gamma^2}{a^4+\gamma^2\Omega^2} + {\cal O}(\epsilon^2),
\end{eqnarray}
where we replaced $\frac{2NT\Omega }{a^2 - \Omega ^2}$ by $L$ using \eqref{EL-from-TO}.

The reason for non-zero $\la xy \ra$ is much more apparent in the rotating frame. Recall the Langevin equation \eqref{langevin-rot} from \ref{app-derivation} which models the dynamics in the rotating frame of reference:
\begin{eqnarray}
    \tfrac{\d}{\d t} \r' &=& \v',
    \qquad
    \\
    \tfrac{\d}{\d t} \v' &=& 
    -\nabla_{\r'} V(\R_t\r')
    -2\, \o \times \v'
    -\o \times( \o \times \r' )
    \nonumber
    \\
    &&-\gamma \v'
    +\sqrt{2\gamma T} \;\bxi(t) .
\end{eqnarray}
Note that in the case of symmetric potential, we have $ V(\R_t\r')=V(\r')$. Otherwise, it acts as a time-dependent driving force in the system: imagine an elliptic potential rotating clockwise with angular velocity $\Omega$. First, consider the case when the potential does not rotate. The steady-state distribution over $\r$ has the form $\rho(\r')\propto\ee^{-\beta V(\r')}$, i.e. the same elliptical shape. 
Now, when external potential rotates and drives the system, the steady-state distribution tries to ``catch up'' the distribution $\rho_t(\r')\propto\ee^{-\beta V(\R_t\r')}$, but will always fall behind. If we go back to the rest frame, the $V(\r)$ is time independent and the steady-state distribution will be slightly twisted in the direction of the rotation.

\section{Derivation of autocorrelation functions}\label{app-autocor}
With symmetric quadratic potential \eqref{potential-sym}, dynamics in $z$ direction completely factorize from the $x$ and $y$ directions in Langevin equation \eqref{langevin}. This is also apparent from the results in \ref{app-FP-OU}, see equations (\ref{sigma-rr},\ref{sigma-vv},\ref{sigma-rv}). Therefore, in this section, we only consider the dynamics of the $x$ and $y$ components. Symmetry of the potential allows us to write the Langevin equation \eqref{langevin} in complex variables:
\begin{eqnarray}
    \ddot{\cz} = 
    -a^2 \,\cz
    -\gamma (\dot{\cz}-\ii \Omega w)
    +\sqrt{2\gamma T} ~\xi(t),
\end{eqnarray}
where $\cz=x+\ii y$ and $\xi(t)=\xi_x(t)+\ii \xi_y(t)$. We take a Fourier transform of the above equation and solve for $\tilde{\cz}(\omega)$:
\begin{equation}\label{complex-langevin-ft}
    \tilde{\cz}(\omega)=\frac{\sqrt{2\gamma T}}{a^2+\ii\gamma(\omega-\Omega)-\omega^2}~\tilde{\xi}_z(\omega),
\end{equation}
where we used the following transformation rules:
\begin{equation}
    \txi(\omega) = \int_{-\infty}^{\infty} e^{- i \omega t} \xi(t) \d t ,
    \qquad 
    \xi(t) =\frac{1}{2\pi} \int_{-\infty}^{\infty} e^{i \omega t} \txi(\omega)
\d \omega.
\end{equation}
Define the correlation function for $\cz$ as 
\begin{equation}
    C_\cz(t) = \la \cz(t)\,\cz(0)^*\ra,
\end{equation}
where the asterisk ($*$) denotes complex conjugation.
According to the Wiener-Khinchin theorem
\begin{equation}
    \la \cz(\omega)\,\cz(\omega')^\ast \ra=2\pi\,\delta(\omega-\omega')\,S(\omega) ,
\end{equation}
where $S(\omega)$ is the spectral density, which is the Fourier transform of the correlation function $C_\cz(t)$. Applying this to the white Gaussian noise, we get
\begin{eqnarray}\label{white-spectr}
    \la \txi(\omega)\,\txi(\omega')^\ast \ra = & 2\cdot2\pi\,\delta(\omega-\omega').
\end{eqnarray}
Factor 2 comes from independent $x$ and $y$ components of the noise. Combining the above with \eqref{complex-langevin-ft}, we get
\begin{equation}
    S(\omega) = \frac{4\gamma T}{\left(a^2-\omega^2\right)^2+\gamma^2(\omega-\Omega)^2},
\end{equation}
and the autocorrelation function is related by the inverse Fourier transform:
\begin{equation}
    C_\cz(t)=\frac{1}{2\pi} \int_{-\infty}^{\infty} e^{i \omega t} S(\omega) \d
\omega.
\end{equation}
We evaluate the integral by contour integration and residue theorem \cite{visual-complex-analysis}. The integrand has the following poles:
\begin{eqnarray}\label{pole-line-1}
    \frac{1}{2} \left(+\ii \gamma -\sqrt{4 a^2-\gamma  (\gamma +4 \ii \Omega )}\right),
    \quad
    \frac{1}{2} \left(+\ii\gamma+\sqrt{4 a^2-\gamma  (\gamma +4 \ii \Omega )}\right),
    \\
    \frac{1}{2} \left(-\ii \gamma-\sqrt{4 a^2-\gamma  (\gamma -4 \ii \Omega )}\right),
    \quad
    \frac{1}{2} \left(-\ii \gamma +\sqrt{4 a^2-\gamma  (\gamma -4 \ii \Omega )}\right),
    \label{pole-line-2}
\end{eqnarray}
For $t>0$, the contour is closed above the real axis, and only the poles with a positive imaginary part will contribute to the integral. Therefore, the integral equals
\begin{eqnarray}\label{form-of-Cz}
    C_\cz(t) = \ii \,\text{Res}(S(\omega), \omega_1) \ee^{\ii \omega_1 t}+\ii\; \text{Res}(S(\omega), \omega_2) \ee^{\ii \omega_2 t}
\end{eqnarray}
where $\omega_1$ and $\omega_2$ are the poles with positive imaginary part when $t>0$, and $\text{Res}(S(\omega), \omega_1)$ is the residue of the function $S(\omega)$ at point $\omega_1$.
All possible autocorrelation functions of processes $x(t)$ and $y(t)$ are present in $C_\cz(t)$:
\begin{eqnarray}
    ~\quad C_{xx}(t) = C_{yy}(t) &&= \frac{1}{2}\Re(C_\cz(t)),
    \\
    \la x(t) y(0)\ra =C_{xy}(t) &&=  -\frac{1}{2}\Im(C_\cz(t)),
    \\
    \la y(t) x(0)\ra =C_{yx}(t) &&=  +\frac{1}{2}\Im(C_\cz(t)),
\end{eqnarray}
where $\Re(~)$ and $\Im(~)$ are real and imaginary parts respectively.

Let $h(t)=x(t)y(t)-\la x y\ra$. Its autocorrelation function is defined as
\begin{equation}\label{Ch-def}
    C_h(t)=\la h(t) h(0)\ra = \la x(t)y(t)x(0)y(0)\ra - \la x y\ra^2.
\end{equation}
Recall the following property of the higher-order moments of the Gaussian (also called Wick's theorem):
\begin{eqnarray}\label{wick}
    E \left[ x_1 x_2 x_3 x_4\right]=&E\left[ x_1 x_2\right]E\left[ x_3 x_4\right]
    +E\left[ x_1 x_3\right] E \left[ x_2 x_4\right]
    \\
    &+E\left[x_1 x_4\right] \nonumber
E\left[x_2 x_3 \right].
\end{eqnarray}
Applying this to \eqref{Ch-def}:\begin{eqnarray}
    C_h(t)=& C_{xx}(t)C_{yy}(t)+C_{xy}(t)C_{yx}(t)
    = \frac{1}{4}\Re\left(C_\cz(t)^2\right)
    \label{autoc-h}
\end{eqnarray}
As $C_\cz(t)$ takes the form of \eqref{form-of-Cz}, the following expressions are easy to evaluate:
\begin{eqnarray}\label{Ch-calc}
    C_h(0)=\!\frac{T^2}{\left(a^2-\Omega ^2\right)^2} 
    \quad\text{and}\quad 
    \int_0^\infty \!\!\!C_h(t)\d t =& \frac{T^2 (a^2+\gamma^2)}{2 \gamma  (a^2-\Omega^2 ) (a^4+\gamma^2\Omega^2)}.
\end{eqnarray}
From expression of $C_\cz$ in \eqref{form-of-Cz} and poles in (\ref{pole-line-1}, \ref{pole-line-2}), we deduce the behaviour of the autocorrelation function $C_h(t)$: for small values of $\gamma$,
$C_h(t)$ oscillates with frequency $2a$ and decays with rate
\begin{eqnarray}\label{Ch-decay-rate}
    \gamma\left(1- \frac{\Omega}{a}\right)
\end{eqnarray}

\section{Estimators for Temperature and Angular velocity}\label{app-TO}

\begin{figure}
    \centering
    \includegraphics[width=0.6\linewidth]{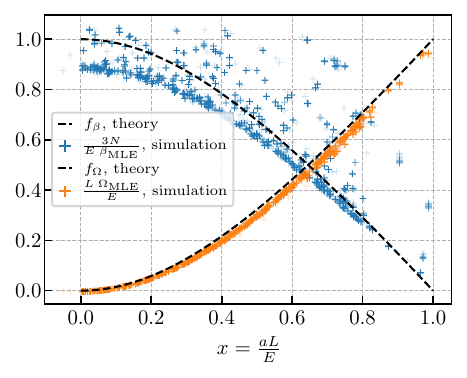}
    \caption{Same as Figure \ref{fig-compare-1}, but a 1000 times more dense system is considered.
    We now see that the numerical estimates of functions $f_\beta$ and $f_\Omega$ are systematically lower than the corresponding dashed lines.
    }
    \label{fig-compare-dense}
\end{figure}

\begin{figure}
  \centering
  \includegraphics[width=\linewidth,trim={0.3cm 0.3cm 0.2cm 0.25cm},clip]{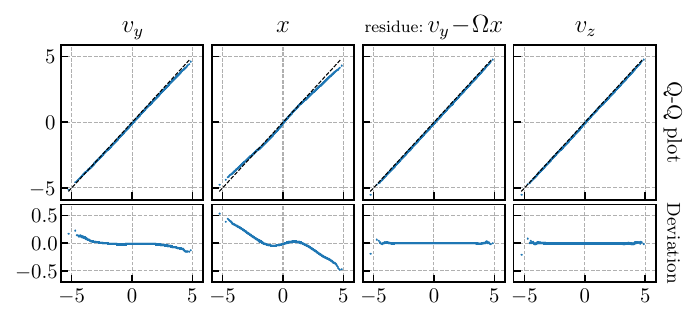}
  \caption{
    \textbf{(top row)} Quantile-quantile plots (Q-Q plot, \cite{thode2002testing})  of 
    $x$ component of the position,
    $y$ component of the velocity and the residue defined in \eqref{residue}, {$z$ component} of the velocity.
  \textbf{(bottom row)} Deviation from the slope (dashed line). 
  In the case of exact Gaussian distributed values, the points in the top row should perfectly align with the dashed line, and deviations should be zero in the bottom row. 
  We conclude that the distribution of residues \eqref{residue} and the $z$ component of velocity are the closest to the Gaussian. 
  This graphical approach surpasses statistical tests, such as Pearson's $\chi$-squared test, as it allows for direct visualization of deviations. Importantly, this method is particularly useful when dealing with correlated data, which is the case for our data. }
  \label{fig-QQ}
\end{figure}

The canonical distribution \eqref{canonic-2} is not Gaussian for the general confining potential $V(\r)$. In this section, we aim to derive potential independent estimators for temperature $T$ and angular velocity $\Omega$.
When the system is not dilute enough (density $\rho\lessapprox\sigma_0^{-3}$, see Section \ref{sec-high-dens}) the interparticle interactions create an effective mean-field potential for a single particle.
Therefore, 
even with quadratic external potential, the potential mean force that a single particle experiences is anharmonic. As we also investigate the high-density regime in Section \ref{sec-high-dens}, we need estimators for the temperature and angular velocity, which will work with the arbitrary potential $V(\r)$.

The non-quadratic potential $V(\r)$ in canonical distribution \eqref{canonic-2} results in a non-Gaussian distribution for position $\r$. However, $\r$ and velocity $\v$ are coupled through the angular velocity $\bs{\Omega}$ (see \eqref{canonic-2}), and this coupling affects the distribution of $\v$ as well, see quantile-quantile plots in Figure \ref{fig-QQ}. We subtract this coupling from the velocity and define a residue term
\begin{eqnarray}\label{residue}
\v_{\text{res}} = \v - \o \times \r.
\end{eqnarray}
which has Gaussian distribution (to see this, change variables from $(\v,\r)$ to $(\v_{\text{res}}, \r)$ and \eqref{canonic-2} will be factorized), which is recorded in the Q-Q plot too, see Figure \ref{fig-QQ}.
Note that $v_z = (\v_{\text{res}})_z$, which means $v_z$ is also Gaussian distributed. Indeed, $v_z$ is independent from the position in canonical distribution $\rho_\text{c}(\r,\v)$, see \eqref{canonic-2}.

To derive potential independent estimators, instead of joint distribution $\rho_\text{c}(\r,\v)$ we  consider the conditional distribution ${\rho_\text{c}(\v\mid \r)}$, which is independent of $V(\r)$:
\begin{eqnarray}\label{v-conditional-x}
    \rho_\text{c}(\v\;|\;\r)&=\frac{\rho_\text{c}(\v,\,\r)}{\int \rho_\text{c}(\v,\,\r)\,\d\v}
    \\
    &\propto
    \exp{\left[-\beta\left(\frac{1}{2}\v^2- \Omega (x v_y-y v_x)
    +\frac{\Omega^2}{2}(x^2+y^2)\right)\right]},
\end{eqnarray}
where $z$-axis is along $\o$ and $\o=\hat{\bs{z}}\Omega$.
Define the maximum likelihood estimators (MLE) of $\Omega$ and inverse temperature $\beta$ as 
\begin{eqnarray}
    (\hat{\beta}_\text{MLE},\;\hat{\Omega}_\text{MLE}) &= \text{arg}\max_{\beta,\;\Omega} \prod_i \rho_\text{c}(\v_i\;|\;\r_i),
\end{eqnarray}
where $(\r_i, \v_i)$ are the position and velocity of a single particle sampled from the stationary distribution $\rho_\text{c}(\v\;|\;\r)$.
After optimizing the likelihood function, we get
\begin{eqnarray}
    \label{MLE-1}
    \hat{\Omega}_\text{MLE} &= \frac{\la x v_{y}-y v_{x}\ra}{\la x^2 + y^2 \ra},
    \\
    \left.\hat{\beta}_\text{MLE}\right.^{-1} \!\!&=\frac{1}{3}\left[
    \la (v_x+\hat{\Omega}_\text{MLE}\; y)^2\ra
    +
    \la (v_y-\hat{\Omega}_\text{MLE}\; x)^2\ra
    +
    \la v_z^2\ra
    \right] 
    \label{MLE-middle}
    \\
    &=\frac{1}{3}\la |\v_\text{res}|^2 \ra,\label{MLE-last}
\end{eqnarray}
where $\la~\ra$ is the empirical average from the sampled points.
In practice,  $(\r_i, \v_i)_{i=1}^{N}$ corresponds to particles in the system.

\section{Fitting of the friction constant $\gamma$: \FP{} for microcanonical ensemble}\label{app-gammafit}

In Section \ref{sec-numerics} we presented three methods of estimation of the friction constant $\gamma$. In this section, we elaborate on the second method.

We have already discussed the \FP{} equation for general Ornstein-Uhlenbeck processes \eqref{FP-OU} and stated its full solution in \eqref{FP-full-sol}. In differential form, it will read
\begin{eqnarray}
    \frac{\d}{\d t}\Sigma(t) = -\Upsilon\, \Sigma(t) - \Sigma(t)\, \Upsilon^\TT +2\rD. \label{sigma-dyn}
\end{eqnarray}
In our case, $\bs\mu(t)=\bs\mu(0)=0$ and \FP{} coefficients are given by \eqref{FP-coefs}.

Consider the time derivative of the average energy:
\begin{eqnarray}
    \label{dE-dt}
    \frac{\d}{\d t}\la E_1 \ra &=\frac{\d}{\d t}\left(\la \frac{1}{2}v^2\ra + \frac{1}{2} \la \r^\TT {\rm A} \r\ra \right)
    \\
    &= \frac{\d}{\d t} \left(\frac{1}{2}\Sigma_{v_x v_x}+\frac{1}{2}\Sigma_{v_y v_y} + \dots\right)
    \\
    &=\gamma (3 T - \la v^2 \ra + L_1 \;\Omega),
\end{eqnarray}
where $L_1=\la x v_y -y v_x \ra$ is the angular momentum of a particle, $v=|\v|$, and we replaced the time derivatives of $\Sigma$ with \eqref{sigma-dyn}. In general, \eqref{dE-dt} is not zero.
The total energy of the system is $E=E_1+\ldots+E_N$ where $E_i$ is the energy of the $i$-th particle. We expect total energy to be constant: $E=\la E \ra = \la E_1 \ra + \ldots + \la E_N \ra$. Thus, the right-hand side of \eqref{dE-dt} must vanish. It cannot be achieved by $\gamma=0$, and therefore we must have
\begin{eqnarray}
    T(\Sigma)&=&\frac{1}{3}\left(\la v^2 \ra - L_1 \;\Omega \right)
    \\
    &=&\frac{1}{3}\left( \Sigma_{v_xv_x}+\Sigma_{v_yv_y}+\Sigma_{v_zv_z} - \Omega (\Sigma_{xv_y}-\Sigma_{yv_x})\right),
    \label{T-relation}
\end{eqnarray}
We interpret this result as follows: 
the temperature serves as an indicator of the magnitude of fluctuations within the system. Fluctuations drive the system: when the fluctuations are high, energy is injected into the subsystem from the thermostat (in this case the thermostat represents the rest of the system). Conversely, when fluctuations are weak, energy is transferred to the thermostat through dissipative processes.
Therefore, the temperature is chosen to ensure that there is no net energy transfer between the subsystem and the remainder of the system.

Now consider the time derivative of $L_1$:
\begin{eqnarray}
\frac{\d}{\d t}L_1 &=& \frac{\d}{\d t}\left(\Sigma_{xv_y}-\Sigma_{yv_x}\right) 
\\
&=& 4\epsilon a^2 \la xy \ra -\gamma \left(L - \Omega \la x^2 + y^2\ra \right).
\label{dL1}
\end{eqnarray}
To be consistent with the angular momentum-torque balance equation \eqref{dL}, the second term in \eqref{dL1} must be zero at all times:
\begin{eqnarray}
    \Omega(\Sigma) = \frac{\la x v_y - y v_x\ra}{\la x^2+y^2\ra } = \frac{\Sigma_{xv_y}-\Sigma_{yv_x}}{\Sigma_{xx}+\Sigma_{xy}}.
    \label{O-relation}
\end{eqnarray}
Surprisingly, the relations in \eqref{T-relation} and \eqref{O-relation} exactly match with our maximum likelihood estimators in (\ref{MLE-1},\ref{MLE-middle}).

So we establish the state-dependent temperature $ T(\Sigma)$ and angular velocity $\Omega(\Sigma)$. They appear in the definitions of $\Upsilon$ and $\rD$ in \FP{}  equation \eqref{FP-OU} and also in \eqref{sigma-dyn} (recall \eqref{FP-coefs} for definitions), thus, the \FP{} coefficients are also state-dependent. Now, there is no closed-form solution for $\Sigma(t)$, but the following numerical integration scheme can be applied:
\begin{eqnarray}
    \Sigma(t+\delta t) =& \Sigma(t) + (-\Upsilon\, \Sigma(t) - \Sigma(t)\, \Upsilon^\TT +2\rD) \; \delta t,
    \\
    T(t+\delta t) =& \left. \frac{1}{3}\left( \Sigma_{v_xv_x}+\Sigma_{v_yv_y}+\Sigma_{v_zv_z} - \Omega (\Sigma_{xv_y}-\Sigma_{yv_x})\right)\right|_{t+\delta t},
    \\
    \Omega(t+\delta t)=& \left. \frac{\Sigma_{xv_y}-\Sigma_{yv_x}}{\Sigma_{xx}+\Sigma_{xy}} \right|_{t+\delta t},
    \\
    \gamma(t+\delta t) =& \gamma(T(t+\delta t)),
\end{eqnarray}
where $\delta t$ is the time discretization. Note that we allowed the temperature-dependent friction constant $\gamma(T)$. In practice, we assume time-dependent $\gamma(t)$ and fit it subject to $N L_1(t)$ matching to the observed $L(t)$ from the simulations. Note that $L_1(t)=\la x v_y -y v_x \ra = \Sigma_{xv_y}(t)-\Sigma_{yv_x}(t)$.

\section{Averaging with respect to microcanonical distribution}
\label{app-micro-avg}
In this section, we show that the average of $x_iy_i$ with respect to microcanonical distribution 
\begin{eqnarray}
    \rho_\text{m}(\psv)=\mathcal{N}\,\delta\left(E-H(\psv)\right)\delta\left(L-L(\psv)\right)
\end{eqnarray} 
is zero, even when the external potential is non-isotropic. Here $\mathcal{N}$ is a normalisation constant, $\psv=\{\r_i,\p_i\}_{i=1}^{N}$ denotes phase space variables with position $\r_i=\{x_i,y_i,z_i\}$ and momentum $\p_i=\{p_{x\,i},p_{y\,i},p_{z\,i}\}$,
\begin{eqnarray}
    H(\psv) &=& \sum_i^N \left(\frac{\p_i}{2}+ V(\r_i)\right)+ \sum_{i<j}\mathcal{U}\left(\left|\r_i-\r_j\right|\right)
    \\
    L(\psv)&=& \sum_i^N \left(x_i p_{y\,i}-y_i p_{x\,i}\right)
\end{eqnarray}
Note that the interaction energy only depends on the relative distance between particles. Also, assume the potential $V(\r)$ is symmetric with respect to following reflections:
\begin{eqnarray}
    \text{point~reflection\!:}\qquad &V(\r)&=V(-\r)
    \\
    \text{plane~reflactions\!:} &V(x,y,z)&=V(-x, y, z)
\end{eqnarray}
For example, potential \eqref{potential-perturbed} satisfies these conditions.
The average of $x_1y_1$ reads
\begin{eqnarray}
    \la x_1y_1\ra = \int \d\r\d\p\; \rho_\text{m}(\r,\p) \; x_1 y_1
\end{eqnarray}
Now, we make the following change of variables:
\begin{eqnarray}
    x_i &\rightarrow -x_i
    \\
    p_{y\,i} &\rightarrow -p_{y\,i} \qquad i=1,\dots,N
\end{eqnarray}
and note that the microcanonical distribution $\rho_\text{m}$ is invariant with respect to this change:
\begin{eqnarray}
    \rho_\text{m}\left(
    \left\{
x_i,
y_i,
z_i,
p_{x\,i},
p_{y\,i},
p_{z\,i}\right\}_{i=1}^N
    \right)
    =
    \rho_\text{m}\left(
    \left\{-x_i,
y_i,
z_i,
p_{x\,i},
-p_{y\,i},
p_{z\,i}\right\}_{i=1}^N    \right)
\end{eqnarray}
while $x_1y_1$ transforms to $-x_1y_1$. Therefore the average vanishes:
\begin{eqnarray}
    \la x_i y_i\ra = 0 \qquad i=1,\dots,N \;.
\end{eqnarray}

\section{Space-dependent friction coefficient}
\label{app-space-gamma}
The density of the system decreases away from the center, see \eqref{rho-x}. Therefore the collisions are less frequent and the friction coefficient is smaller compared to the value at the center. It is also evident from Figure \ref{fig-snap} that the outer trajectories are in ballistic motion.

We still use \FP{} equation \eqref{FP0} with $\gamma(\r)$ now function of position. 
In the case of symmetric potential, \eqref{canonic} is again the steady-state solution, even though $\gamma$ now depends on $\r$. When the external potential is quadratic, steady-state distribution \eqref{canonic} is Gaussian, and this property will be used below.

Now we parametrize $\gamma(\r)$ as  
\begin{eqnarray}
    \gamma(\r) = \gamma_0+\gamma_1 (x^2+y^2)
\end{eqnarray}
where the behaviour of $\gamma$ presented in the begining of the section is accounted for by $\gamma_1<0$.

Now we find the quasi-stationary distribution of this new \FP{} equation by the method of moments. We multiply the \FP{} by all possible pairs of variables (e.q. $x^2$, $x v_y$, etc.) and perform integration by parts. This results in equations for moments. For example, if we multiply by $x^2$ we will get
\begin{eqnarray}
    0&=& 
    -\int \d\r\d\v\, x^2 \left(\partial_x (v_x P)+\partial_y (v_y P) + \partial_z (v_z P) \right)
    \\
    &&+\!\!\int \d\r\d\v\, x^2\nabla_{\v}  \left[ \left(\nabla_{\r} V(\r)
    +\gamma(\r) (\v-\o\times\r) \right) P \right]
    + \!\!\int \d\r\d\v\,x^2\gamma T \,\nabla_{\v}^2 P \nonumber
    \\
    &=&+\int \d\r\d\v\,2 x v_x P + 0 + 0
    \\
    &=& 2\la x v_x \ra 
\end{eqnarray}
or, if we multiply by $x v_x$ we get
\begin{eqnarray}
    0&=& 
    -\int \d\r\d\v\, x v_x \nabla_{\r} \left[ \v P \right]
    \\
    &&+\int \d\r\d\v\, xv_x\nabla_{\v}  \left[ \left(\nabla_{\r} V(\r)
    +(\gamma_0+\gamma_1(x^2+y^2)) (\v-\o\times\r) \right) P \right] \nonumber
    \\
    &&+ \int \d\r\d\v\, xv_x\gamma T \,\nabla_{\v}^2 P,
    \\
    &=& \la v_x^2\ra - a^2 (1+\epsilon)^2 \la x^2\ra 
    \gamma_0\left(\la xv_x\ra + \Omega \la xy\ra\right)
    \\
    &&-\gamma_1 \left(\la x^3v_x\ra + \la xy^2v_x\ra +\Omega\la x^3y\ra+\Omega\la xy^3\ra 
    \right)
\end{eqnarray}
where we already see the higher-order moments. As the quasi-stationary distribution $P$ is a weakly perturbed version of Gaussian distribution \eqref{canonic}, we approximate higher order moments by Wick's theorem (see \eqref{wick}). Eventually, we get a closed system of equations for 2nd order moments.

In the limit of small $\gamma_0$ and $\gamma_1$ we have
\begin{eqnarray}
    \la xy\ra = \frac{2 T \Omega  \epsilon }{\left(a^2-\Omega^2\right) \left(\gamma_0+\frac{2 \gamma_1 T}{a^2}\frac{a^2+\Omega ^2}{a^2-\Omega ^2}\right)}
\end{eqnarray}
and in comparison with \eqref{xy-mean-app}, we infer the effective single parameter $\gamma$:
\begin{eqnarray}
    \gamma = \gamma_0+\frac{2 \gamma_1 T}{a^2}\frac{a^2+\Omega ^2}{a^2-\Omega ^2}
\end{eqnarray}

\section{Absence of collisionless relaxation}
\label{app-collisionless}
There are many examples in physics (mainly in quantum mechanics) where collisionless relaxation is the main mechanism of the mixing and relaxation of the system \cite{goldman2001quantum}. However, this is not the case for our system when we turn off the interactions. Each of the particles will just oscillate independently:
\begin{eqnarray} \label{h1}
    x(t)=x_0 \cos\left(a(1+\epsilon) t\right)+\frac{v_{x0}}{a(1+\epsilon) } \sin\left(a(1+\epsilon) t\right),
    \\
    \label{h2}
    y(t)=y_0 \cos\left(a(1-\epsilon) t\right)+\frac{v_{y0}}{a(1-\epsilon) } \sin\left(a(1-\epsilon) t\right),
\end{eqnarray}
and particles are different from each other only in initial positions and velocities $x_0, y_0, v_{x0}, v_{y0}$. We write the total angular momentum as 
\begin{eqnarray}
    L(t) = \sum_{i=1}^{N} \big(x_i(t) \dot{y}_i(t)- y_i(t) \dot{x}_i(t) \big)
\end{eqnarray}
where $x_i(t)$ and $y_i(t)$ are \eqref{h1} and \eqref{h2} with different (and independent) initial conditions. The average of $L(t)$ with respect to initial canonical distribution \eqref{canonic} is
\begin{eqnarray}
    \la L(t) \ra &=& \la L(0) \ra \left(\frac{1}{1-\epsilon^2}\cos(2a\epsilon t) -\frac{\epsilon^2}{1-\epsilon^2}\cos(2 a t) \right)
    \\
    &\approx&\la L(0) \ra \; \cos(2a\epsilon \;t)
    \label{L-cos-decay}
\end{eqnarray}
for small $\epsilon\ll1$. Even though \eqref{L-cos-decay} possesses an initial decay of $L$, it does not describe a relaxation process, $\la L(t)\ra$ is periodic and the period does not increase by the number of degrees of freedom. After time $\frac{\pi}{a\epsilon}$, the total angular momentum returns to the initial value.

\section{Details of numerical simulations}\label{app-numerics}

For numerical simulations, we have used LAMMPS \cite{LAMMPS}, code is available in the public GitHub repository \cite{ashot-github}. Lennard-Jones potential was used with a cutoff $5\sigma_0$.
We take unit particle masses $m=1$, and unit Boltzmann constant $k_B=1$. Time, distance and energy are dimensionless quantities.

\subsection{Random initialization with fixed E and L}
\label{app-random-EL}
In our investigation of ensemble behavior, it is essential to 
initialize the system with exactly the same fixed energy $E$ and angular momentum $L$. 
Suppose we want to generate random $\r_i$ and $\v_i$ such that the total energy and
total angular momentum are the given $E$ and $L$. 
The initialization algorithm proceeds as follows:
\begin{enumerate}
    \item Place particles on a grid, such as a Face-Centered Cubic
lattice. Put particle on lattice site $\r$ if $V(\r) < E_1$. Suppose there are $N$ particles satisfying this condition. 
\end{enumerate}
In practice, we specify the target energy per particle $E_1$ and the total energy becomes $E=N\, E_1$. We control the number of particles by adjusting either the lattice side length or the stiffness of the external potential.
Let $I=\sum_{i=1}^N \left( x_i^2 + y_i^2 \right)$ represent the
moment of inertia along the $z$ axis. 
\begin{enumerate}
    \setcounter{enumi}{1}
    \item initialize velocities $\v_i \overset{\text{i.i.d.}}{\sim} \mathcal{N}(0,1)$ from standard
normal distribution independently.
\end{enumerate}
Let
$E_\text{init}=\sum_{i=1}^N \frac{1}{2}\v_i^2$ and $L_\text{init}=\sum_{i=1}^N
(\r_i \times \v_i)_z$ with the above velocities.
Set $E_\text{target}=E-V_\text{interactions}(\{\r_i\}) -
V_\text{external}(\{\r_i\})$ as the target kinetic energy. 
Consider the following transformation:
\begin{eqnarray} \label{velocity-transform}
    \v'_i = \alpha \v_i + \ozg \,\ez\times\r_i,
\end{eqnarray}
where $\alpha$ and $\ozg$ are parameters chosen to ensure that the total kinetic energy equals $E_\text{target}$ and the total angular momentum equals $L_\text{target}$. Note that
\begin{eqnarray}
    \qquad{\v'_i}^{2} &=& \alpha^2 \v_i^2 + \ozg^2 (x_i^2+y_i^2) +  \alpha\ozg
(\r_i \times \v_i)_z,
    \\
    \r_i \times \v_i' &=& \alpha\, \r_i \times \v_i + \ozg \,\ez (x_i^2+y_i^2),
\end{eqnarray}
and therefore
\begin{eqnarray}
    E_\text{target}=& \,\alpha ^2 E_\text{init}+\tfrac{1}{2}I \ozg ^2+\alpha\,
L_\text{init} \ozg 
    \\
    L_\text{target}=& \,\alpha L_\text{init} + I \ozg .
\end{eqnarray}
We solve for $\alpha$ and $\ozg$ and select the solution with a positive $\alpha$:
\begin{eqnarray}\label{transform-sol}
    \alpha =& \sqrt\frac{{2I E_\text{target}-L_\text{target}^2}}{{2 I E_\text{init}-L_\text{init}^2}},
    \qquad
    \ozg = \frac{L_\text{target}-\alpha L_\text{init}}{I}
\end{eqnarray}
It is important to note that for a fixed moment of inertia $I$, i.e. for a fixed lattice of particles, the following condition must be satisfied:
\begin{equation}\label{Lmax}
    L^2 \le 2 I E_\text{kinetic}\;.
\end{equation}
This implies that $L$ cannot exceed a certain limit for a given energy.
And the final step of the algorithm:
\begin{enumerate}
    \setcounter{enumi}{2}
    \item transform velocities by the relation \eqref{velocity-transform} with
solution \eqref{transform-sol}.
\end{enumerate}
The randomness of the initial state only comes from the randomness of initial velocities.

\subsection{Parameters of the simulations}
In this section, we present the parameters used in our simulations. Throughout our simulations, we employed Lennard-Jones potential defined in \eqref{lennard}
with $\varepsilon=0.1$ and $\sigma_0=0.05$.
In order to reduce the system's size in the $z$-direction while maintaining a three-dimensional system, we consistently set $a_z=4\cdot a$ in the external potential  \eqref{potential-perturbed}.

For the results depicted in Figure \ref{fig-compare-1}
we varied the energy per particle $E_1$ in the range $0.1-20$, and the potential parameter $a$ in the range $0.03-1.6$. We keep the initialization lattice side length fixed: $\sigma=1\gg\sigma_0$. This assures low system density. 
Each pair of $(E_1, a)$ will correspond to some number of particles $N$ after initialization. We considered pairs only when N fell within the range of 10 to 600. Furthermore, we choose the value of initial angular momentum $L$ uniformly random between $0$ and the maximum value allowed by \eqref{Lmax}. 
We simulate with the time discretization $\delta t =10^{-4}$ and total simulation time $t=2\cdot 10^4$ and we use the last 30\% of the data for estimations.

For the rest of simulations we have used $\delta t = 10^{-3}$, $a=0.2$, $E_1=1$ and $N=523$, i.e. the total energy of the system is $E=523$. With these parameters, $\gamma$ is on the order of $10^{-3}$.
To establish a microcanonical distribution with a given $E$ and $L$ and to eliminate any correlations with the initial state, we always simulate the system in the symmetric potential for $5000$ units of time before breaking the symmetry.

For the simulations presented in Figure \ref{fig-L} we have taken $\epsilon=3.6\cdot10^{-4}$ and run $1200$ independent simulations with identical initial $L$ for $7\cdot 10^5$ units of time. The objective was to ensure that $L$ decays less than 1\% in each interval of size $\gamma^{-1}\sim 10^3$ so that $L$ were a slow variable and the system was in a quasi-stationary state throughout the simulation.

Figure \ref{fig-snap} visualizes a 5-unit-time interval dynamics of the system of Figure \ref{fig-L} at $L\simeq1500$, i.e. small part of one simulation out of 1200 independent simulations.

Figure \ref{fig-gamma} is based on the data of Figure \ref{fig-L} and the analysis methods are detailed in Section \ref{sec-mean-relax}.

In Figure \ref{fig-gamma-T} we only altered $\epsilon$ in simulations compared to the simulations of Figure \ref{fig-L}. Likewise, 1200 independent simulations were used for each $\epsilon$.

For Figure \ref{fig-CS}, simulations started with the same $E$ and $L=0$ and were performed for $2\cdot 10^5$ units of time. Autocorrelation and spectral density were estimated on the last 75\% of the data.

In Figure \ref{fig-compare-dense} we have used lattice side length $\sigma=0.1$, 10 times smaller than in Figure \ref{fig-compare-1}. The rest of the parameters were kept the same. For Figure \ref{fig-QQ}
we considered the subset of simulations with $a=0.5$, $N=523$ and  $L$ the 60\% of maximum possible value.

\end{document}